# Unveiling a new shear stress transfer mechanism in composites with helically wound hierarchical fibres

A. Cutolo[1], A.R. Carotenuto[1], S. Palumbo[1], F. Bosia[2], N. Pugno[3,4] and M. Fraldi[1*]

[1]Department of Structures for Engineering and Architecture, University of Napoli Federico II – Italy
[2]Department of Applied Science and Technology, Politecnico di Torino, Italy
[3]Laboratory of Bio-inspired, Bionic, Nano, MetaMaterials & Mechanics, Department of Civil, Environmental and Mechanical Engineering, University of Trento – Italy
[4]Griffith Theory Centenary Lab, School of Engineering and Materials Science, Queen Mary University of London – UK

**Abstract**

The mechanical performance of reinforced composites is strongly influenced at different scales by the stress transferred at the matrix-fibre interfaces and at any surface where material discontinuity occurs. In particular, the mechanical response of elastomeric composites where the reinforcement is composed by cords with helically wound fibres is heavily compromised by fatigue and delamination phenomena occurring at cord-rubber as well as at the ply interfaces, since rubber and polymeric matrices are mainly vulnerable to the accumulation of deviatoric energy due to the shear stresses transferred across the surfaces. Despite the large diffusion of composites in a vast field of applications and the mature knowledge of their behaviour, some key mechanical aspects underlying failure mechanisms are still partially unclear. For example, stress amplification and strain localization are often difficult to predict by means of analytical solutions and averaging techniques that usually conceal stress gradients. In this work, we analyse coupling between torsional and tensile loads in twisted cords, which are adopted in many cases to reinforce composites and rubbers in tire applications. We provide a model characterized by an enriched cord-matrix mechanical interplay able to theoretically explain and predict actual stress distributions responsible for the onset of delamination and fatigue-guided phenomena that are experimentally observed in these composites. In particular, we demonstrate that the assumption of a monoclinic/trigonal behaviour for the mechanical response of the hierarchical strands allows to estimate, by means of analytical formulas and a homogenization approach, hitherto neglected shear stresses at the matrix-reinforcement interface. These stresses are transferred to the neighbouring regions, leading to post-elastic behaviour and failure events. Finite Element results for rubber/cord composites under various boundary conditions highlight the discrepancies in mechanical behaviour between the proposed model and standard approaches in the literature, while experimental tests performed on composite rubber pads confirm theoretical predictions, providing qualitative evidence of crack initiation.

**Keywords**: Wire strand; Composites; Trigonal materials; Homogenization.

## 1. Introduction

Helicoidal structures are widely employed at all scales in engineering applications, for example for mooring offshore oil platforms, suspended bridges, aircraft landing gear, armoured cables, trusses, ropes and as reinforcements in composites. Also, wound fibres are widely encountered in biomaterials at different scale levels, often in the form of hierarchically organized structures, from protein filaments in cells to bone osteons, muscles and tendons [1, 2], including prestress [3, 4].
A helical arrangement of wires enables to better exploit materials under various loading conditions. A wound structure enhances the wire capacity to support large loads on the helical axis, while bending as well as torsion produce relatively mild states of stress [5]. This peculiar mechanical behaviour can be associated with local relative movements between adjacent wires that produce a coupling between axial and torsional reactions. The act of straining a wire twisted around a core produces both its elongation and the reduction of the angle that the wire forms with the core, resulting in a torque about the strand axis or a twisting in the case of free end conditions. Reaction of the helical wire to elongation directly affects the overall tensile strand stiffness as well as the relative strand resistance. Additionally, the more the wire helical angle changes, the greater the contributions due to wire torsion and bending are. The combination of these conditions determines the global axis response of the strand at any specific state of elongation. Moreover, since bending and torsional stiffness in the strand are directly related to the inertia of the wire cross sections, their values are smaller if the wires are thinner. Inter-wire slippage generally contributes to the system energy dissipation, but it will be neglected in this work, where only conservative effects are considered.

---

* Corresponding author: Massimiliano Fraldi – fraldi@unina.it

Evidently, both intrinsic elastic properties and the arrangement of the wires determine the overall mechanical behaviour of a strand. Quantitative evaluation of the strand response is thus decisive to determine the best design for wound wire structures. For this reason, a large number of experimental tests have been conducted with specific applications in mind [6, 7, 8, 9].

Other authors have instead developed theoretical models for more general cases, thus enabling to predict the global mechanical response of a strand starting from wire material properties and geometrical configurations. Many efforts have been made in the literature to analytically describe the elastic behaviour of strands using bottom-up strategies, starting from the knowledge of wire material properties and structure geometry. A common approach is based on single wire mechanics, which derives the strand response from the kinematics of each wire. In this framework, Hruska [10, 11, 12] and Knapp [13] developed an analytical strand model assuming the hypothesis of a rigid core, McConnell and Zemek [14] then introducing torsional stiffness to improve previous models. The so-called rod mechanics theory, proposed by Love [15], also provides a detailed estimation of the wire behaviour and an accurate description of strands. This helical rod approach was also exploited by Machida and Durelli [16], who introduced an asymmetric stiffness matrix for the strand based on bending and torsional effects. By instead considering the Poisson's ratio effect in wires, Costello and Philips [17] and Velinsky [18] analysed the response of ropes with an internal strand core. Costello [19] then presented a linearized theory able to take into account the effects of twist and curvature variations for any complex arrangement of strands in a rope, by providing an asymmetric stiffness matrix. Based on Costello's cord model, Paris [20] proposed an elasticity approach to analyse the axial force and twisting moment, shear stress transfer and effective modulus of the cord by introducing extension-twisting coupling. Other contributions have been provided by Kumar and Cochran [21], who proposed a closed form expression to compute the maximum contact inter-wire stress. Starting from Love's rod theory, Sathikh et al. [22] modelled a cable with a rigid core using a symmetric linear approach and Labrosse [23] considered wire relative motion omitting the Poisson's ratio effect in a (6+1) wire cable assembly. In the wire rope model by Elata et al. [24], neither torsion nor bending were accounted for, but double-helix behaviour of the strand and inter-wire adjacent load effects were incorporated in the study. Argatov [25] computed the effect of Hertzian contact between the external wires and the core, and Wang et al. [26] improved the model efficiency by computing the helix twisting structure of the wires by means of a mass spring model. An analytical description of the axial-torsional response of single-layer metallic strands providing a symmetric stiffness matrix, in accordance with the Betti-Maxwell reciprocity theorem, was presented by Foti and Martinelli [27], who introduced a relevant extension of the Argatov-based approaches. Further symmetric strand responses were additionally obtained in recent works by Karathanasopoulos et al. [28, 29].

All the mentioned approaches, based on helical rod theory, allow to investigate the influence of the microstructure parameters on the overall strand response, determining global features from geometrical and mechanical characteristics of single wires. Nevertheless, these models are inadequate in terms of strand homogenization procedures, since they are generally unable to treat the strand as a continuum that preserves the intrinsic structural chirality, *de facto* neglecting the actual coupling of twisting and extensional regimes and thus invoking orthotropic symmetries. In the case of multi-layered strands, several authors preferred to choose a semi-continuous approach, where each layer is considered an equivalent orthotropic sheet or cylinder. This strategy is advantageous, especially in the presence of a large number of layers [30, 31, 32, 33], but does not account for the trigonal response of the wire ensemble. Continuous models have also been proposed for biological structures [34] and for ropes made of a large number of wires as well as helically reinforced cylinders [35], finally arriving to standard anisotropic symmetry.

Starting from the well-established stiffness coefficients obtained from the wire microstructure and derivable through the strand model by Argatov [25, 19, 36] within the theoretical framework of naturally curved rods [15], the present work proposes a semi-continuous approach to homogenize the mentioned strand as a cylindrical structure, in which the axial dilation/twisting coupling, due to the orthotropic behaviour of helical wires, can be reproduced in each hollow cylinder by assuming monoclinic symmetry. The solutions of the continuum anisotropic cylinder are obtained analytically following the approach proposed by Lekhnitskii [37], then computing the elastic moduli of the homogenized object directly relating them to the strand geometrical and constitutive parameters. To take into account complex microstructures and hierarchical arrangements of the wires in a strand, the exact solutions have been found by ideally considering Functionally Graded Material Cylinders (FGMCs) made of a central core and an arbitrary number of hollow phases, each one possessing general monoclinic anisotropy in cylindrical coordinates, thus imposing proper continuity conditions on displacements and stresses at the interfaces. Once the closed-form solutions are obtained, we essentially solve a system of algebraic equations to finally establish a direct one-to-one identification of the elastic moduli of the homogenized stiffness matrix of the monoclinic/trigonal composite cylinder with the stiffness coefficients obtained following Argatov [25], thus deriving explicit expressions of the elasticity tensor as a function of wire configuration and geometrical and elastic parameters of the single wires of which the strand is made.

Numerical Finite Element (FE) simulations are then performed to highlight the fact that competition between geometry and mechanics of the strand microstructure – affecting the strand rupture probability [38] – if projected at the continuum level and implemented in much less expensive calculations than those required by computational models describing the wires in detail. Further, this interaction also plays a non negligible role in the mechanics of composites reinforced with helical fibres at any scale, influencing magnitude and distribution of shear stresses across the material interfaces in a relevant manner, producing unexpected stress intensifications, triggering the onset of possible failure phenomena and thus determining the mechanical performance of the composites.

In the final part of the paper, these analytical and numerical predictions are verified through experimental tests on rubber/cord composites under uniaxial tension, thus validating the proposed theoretical approach and its numerical implementation.

## 2. Mechanics of single wires and simple strands

Following the recent work by Fraldi et al. [38], in which the general characterization of hierarchical multi-layer strands was extensively reported, here we investigate the mechanics of a specific structure, namely the simple straight strand constituted by a central core surrounded by one layer of helically wound wires. Let us consider a global coordinate system $\{x_1, x_2, x_3\} \in \mathsf{R}^3$, with $x_3$ as the strand axis (see Figure 1). According to rod theory by Love [15], an additional helical reference system $\{\mathbf{e}_1, \mathbf{e}_2, \mathbf{e}_3\} \in \mathsf{R}^3$ is locally defined over the generic cross section of each wire. In particular, the unit vector $\mathbf{e}_2$ is chosen normal to the bending plane $\mathbf{e}_1 - \mathbf{e}_3$ of the helical wire centreline. Then, a simple straight strand made of (6+1) wires is first adopted to introduce the kinematics of the strand model (see Figure 2) . The strand consists of six wires, all of which with circular cross section whose radius is $R_w$, wrapped helically around a core of radius $R_c$. All the external wires are assumed to lie in the same initial helical configuration, with the helix angle $\alpha$ taken with respect to the strand cross section, as shown in Figure 2. Two further parameters can be introduced, say the complementary angle $\beta = \pi/2 - \alpha$, measured with respect to the strand axis, and the distance of the wire centreline from the strand axis, $r_h = R_c + R_w$. This configuration represents the simplest arrangement of wires in a strand. In order to construct the mechanical model of the strand, some key assumptions must be considered. First, the overall strand length is assumed sufficiently long to avoid end side effects, and a prescribed condition on the radial contact among wires and core is assumed, i.e. the external wires do not touch, but all of them are only in contact with the core. In addition, friction between wires and core is assumed to be sufficiently high to prevent any relative slip. Interlayer pressure effects and contact deformations are instead taken into account.

It is worth highlighting that, focusing on the elastic response of strands, dissipative phenomena such as inter-wire slippage, wire flattening and plasticity are implicitly neglected. In this respect, Utting and Jones [39] specifically focused on inter-wire friction in the case of small deformation of a strand, experimentally showing a small influence of these effects on the global strand behaviour. Nawrocki and Labrosse [40] also performed studies on inter-wire contacts using Finite Element numerical models, and their results demonstrated that rolling and sliding can influence the overall mechanical response of strands through pivoting between external wires and core. However, comparison of results from numerical analyses and experimental data seem to suggest that pivoting is a stress-free phenomenon, at least in the range of small deformations, allowing to neglect these local effects and to simply relate wire kinematics to the overall degrees of freedom of the strand.

As a consequence, according to rod theory, a single wire reacts to overall loads macroscopically applied to the strand exhibiting a combined deformation regime characterized by local elongation, bending and twisting. Moreover, only small deformations are assumed and therefore the equilibrium equations can be referred to the undeformed state. Finally, all the strand filaments are assumed to be isotropic and linearly elastic, with Young's moduli, $E_c$ and $E_w$, and Poisson's ratios, $\nu_c$ and $\nu_w$, for core and wires respectively; equivalently, corresponding Lamé moduli, $\lambda_c$, $\lambda_w$, $\mu_c$ and $\mu_w$, will also be introduced in the following for convenience.

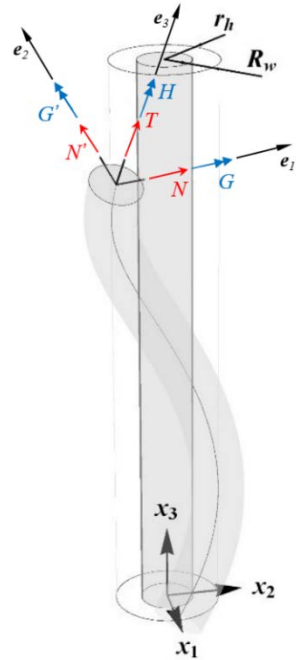

**Figure 1:** Sketch of a generic strand wire arrangement, with global and local coordinate systems, $\{x_1, x_2, x_3\} \in \mathsf{R}^3$ and $\{\mathbf{e}_1, \mathbf{e}_2, \mathbf{e}_3\} \in \mathsf{R}^3$, respectively.

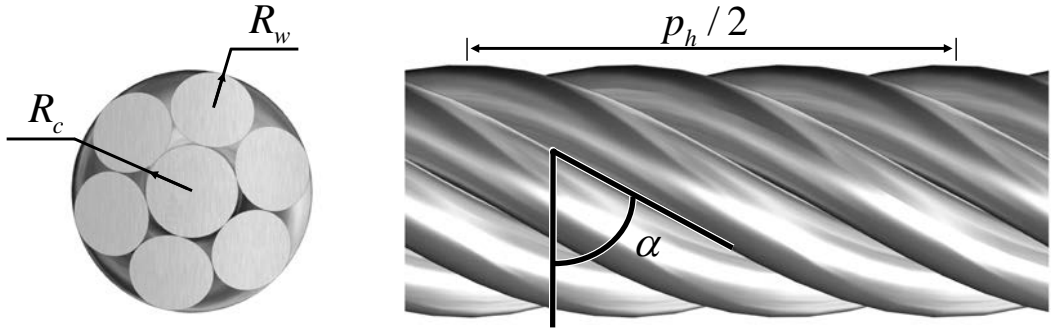

**Figure 2:** Simple (6+1) straight strand: cross section and lateral view.

## *2.1. Kinematics of a simple strand*

In light of the theory of naturally curved rods proposed by Love [15] and following the analytical approach proposed by Argatov [25] to analyse the mechanics of simple straight strands, we derive a kinematical model to predict the response of simple and hierarchical strands subjected to prescribed load or displacement boundary conditions. In particular, self-equilibrated tensile axial forces $F$ and torques $M_t$ exerted at the strand extremities will be considered, together with the corresponding generalized overall deformations, represented by the elongation $\varepsilon$ and the torsion angle per unit length, $\varphi$. For the particular case of a simple strand, the resultant constitutive relations depend on some key geometrical and mechanical parameters: the wire helix angle $\alpha$, the radii of the wire cross sections $R_w$, the core radius $R_c$, the number of external wires $m$, and the Young's moduli and the Poisson's ratios of the constituent materials. Specifically, neglecting overall bending (i.e. the resultant bending moment at the extremities), the deformation of the strand can be described through the overall axial elongation

$$\varepsilon = \frac{L - L_0}{L_0} \tag{1}$$

and the overall twist angle per strand unit length

$$\varphi = \frac{\Delta\Theta}{L_0} \tag{2}$$

where $L_0$ and $L$ refer to strand length in initial and stressed configurations, respectively, and $\Delta\Theta$ denotes the relative twist angle between two strand cross sections at a distance $L_0$. Starting from Ramsey wire rope theory [41, 42], the Argatov model allows to take into account the deformation due to the inter-wire contact by combining both the effect of Poisson's ratio and the asymptotic solution of the plane Hertz's problem, thus including these local phenomena into the overall strand kinematics.

## 2.2. Mechanics of single wires

In the local helical reference system $\{\mathbf{e}_1, \mathbf{e}_2, \mathbf{e}_3\}$ (see Figure 1), the unstrained curvatures of the wires in the strand outer layer can be defined as follows [38]

$$\kappa_{w0} = 0, \quad \kappa'_{w0} = \frac{\cos^2 \alpha}{r_h} \tag{3}$$

where $\kappa_{w0}$ and $\kappa'_{w0}$ are the curvatures in the planes normal to the vectors $\mathbf{e}_1$ and $\mathbf{e}_2$, respectively, while the axial twist around $\mathbf{e}_3$ is

$$\tau_{w0} = \frac{\sin \alpha \cos \alpha}{r_h} \tag{4}$$

As the strand is loaded, each wire deforms by assuming a helical configuration coaxial with the initial helix. The updated curvatures and twist can be expressed as

$$\kappa_{w1} = 0 \quad \kappa'_{w1} = \frac{\cos^2 \alpha_1}{r_{h1}}, \quad \tau_{w1} = \frac{\sin \alpha_1 \cos \alpha_1}{r_{h1}} \tag{5}$$

where $r_{h1}$ and $\alpha_1$ are the radius and the angle of the deformed helix. It is worth noting that, in the absence of overall bending, the curvature $\kappa_{w0}$ in the plane with normal $\mathbf{e}_1$, as well as its variation $\Delta \kappa_w = \kappa_{w1} - \kappa_{w0}$, are vanishing in every section. Hence, the deformation of a wire in the strand can be completely described by the wire elongation measured along its axis, say $\varepsilon_w$, the wire bending curvature in the $\mathbf{e}_1 - \mathbf{e}_3$ plane, $\Delta \kappa'_w = \kappa'_{w1} - \kappa'_{w0}$, and the wire torsion $\Delta \tau_w = \tau_{w1} - \tau_{w0}$. By additionally including the Poisson effect and the Hertzian contact between the core and wires, the radial deformation $\rho$ of the helix radius $r_h$ (the distance between the centre of the core and the centreline of the wires) can be expressed as

$$\rho = 1 - \frac{r_{h1}}{r_h} = \frac{\varepsilon v_c R_c + \varepsilon_w v_w R_w + \delta_r}{R_c + R_w} \tag{6}$$

where $\delta_r$ represents the so-called *contact approach*, a term responsible of the interaction between core and wires [25]. However, as shown in the next paragraph, small strand deformations imply that $\{\varepsilon, \varphi, \rho\} \ll 1$, and consequently second order terms can be neglected in the calculation.

Equilibrium involves the presence of the two shear components directed along $\mathbf{e}_1$ and $\mathbf{e}_2$ directions, say $N$ and $N'$ respectively, and the tensile force, $T$, along $\mathbf{e}_3$ (see Figure 1). Furthermore, three moments emerge on the wire cross sections: two bending moments $G$ and $G'$ in the planes whose normals are the vectors $\mathbf{e}_1$ and $\mathbf{e}_2$, respectively, and the torsion $H$ acting on the cross section plane. Since $\kappa_{w0} = \kappa_{w1} = 0$, the shear force $N$ and the bending moment $G$ are vanishing. Although classical body forces are neglected, the effect of the contact forces per centreline unit length $X$ at the interfaces between the wires and the core is assumed to act as a body force directed along the $\mathbf{e}_1$ direction. Within these hypotheses, the equilibrium equations can be written as

$$\begin{cases} -N'\tau_{w1} + T\kappa'_{w1} + X = 0 \\ -G'\tau_{w1} + H\kappa'_{w1} - N' = 0 \end{cases} \quad (7)$$

Constitutive relations between the nonzero generalized forces and the components of curvature, twist, and elongation are introduced [41]

$$G' = E_w I_w \left( \Delta\kappa'_w + \kappa'_{w0}\, \varepsilon_w \right), \quad H = \mu_w J_w \left( \Delta\tau_w + \tau_{w0}\, \varepsilon_w \right), \quad T = E_w A_w\, \varepsilon_w \quad (8)$$

where $I_w = \pi R_w^4/4$ is the cross-sectional moment of inertia along the $\mathbf{e}_2$ axis, $J_w = \pi R_w^4/2$ is the polar moment of inertia of the wire cross section, $A_w = \pi R_w^2$ represents the wire cross-sectional area, $\mu_w$ and $E_w$ being the shear and the Young moduli of the wire, respectively. The resultant axial force $F^\mathrm{l}$ and twisting moment $M_t^\mathrm{l}$ are then given by

$$\begin{aligned} F^\mathrm{l} &= m\left( T\sin\alpha_1 + N'\cos\alpha_1 \right), \\ M_t^\mathrm{l} &= m\left( H\sin\alpha_1 + G'\cos\alpha_1 + T\, r_{h1}\cos\alpha_1 - N'\, r_{h1}\sin\alpha_1 \right) \end{aligned} \quad (9)$$

where $m$ is the number of wires surrounding the core and l denotes the layer ideally made by wires. Analogously, for the core one has

$$\begin{aligned} F^c &= E_c A_c\, \varepsilon_c, \\ M_t^c &= \mu_c J_c\, \varphi_c, \end{aligned} \quad (10)$$

where $J_c = \pi R_c^4/2$ and $A_c = \pi R_c^2$ are the polar moment of inertia and the area of the core cross section, $\mu_c$ and $E_c$ representing the core Lamè and Young elastic moduli, respectively. For the simple straight strand, the local core strain measures can be taken to be coincident with the global ones, that is $\{\varepsilon_c \equiv \varepsilon,\ \varphi_c \equiv \varphi\}$. Finally, the kinematics of the wires is fully defined through three parameters: the local elongation $\varepsilon_w$ along the wire axis, the local difference in wire curvature $\Delta\kappa'_w$ in the plane whose normal vector is $\mathbf{e}_2$ and the wire twist angle variation $\Delta\tau_w$. In accordance with Costello's formulation [19], the wires belonging to the layer do not touch each other, this condition requiring

$$R_w \sqrt{1 + \frac{\tan^2\left(\dfrac{\pi}{2} - \dfrac{\pi}{m}\right)}{\sin^2\alpha}} < r_h \quad (11)$$

### *2.3. Constitutive equations of the strand*

Constitutive equations for the strand can be derived by expressing the external loads as the sum of the wire forces and moments projected along the strand axis. In this way, the total axial force $F$ and the resultant torque $M_t$ can be related to the generalized strand deformations $\varepsilon$ and $\varphi$ in Eqs. (1) and (2) as follows:

$$\begin{bmatrix} F \\ M_t \end{bmatrix} = \begin{bmatrix} F^c + F^l \\ M_t^c + M_t^l \end{bmatrix} = \begin{bmatrix} k_{\varepsilon\varepsilon}^c + k_{\varepsilon\varepsilon}^l & k_{\varepsilon\varphi}^c + k_{\varepsilon\varphi}^l \\ k_{\varphi\varepsilon}^c + k_{\varphi\varepsilon}^l & k_{\varphi\varphi}^c + k_{\varphi\varphi}^l \end{bmatrix} \begin{bmatrix} \varepsilon \\ \varphi \end{bmatrix}, \quad (12)$$

where the coefficients in the $(2 \times 2)$ matrix represent the strand stiffness constants, whose explicit expression can be derived from Eqs. (9) and (10)

$$\begin{aligned}
k_{\varepsilon\varepsilon}^c &= E_c A_c, & k_{\varepsilon\varepsilon}^l &= m E_w A_w \left[ \sin^3 \alpha + q_2 - (q_1 - q_2) \Lambda_1 \right] \\
k_{\varepsilon\varphi}^c &= 0, & k_{\varepsilon\varphi}^l &= m E_w A_w \left[ (q_1 + q_3)(R_c + R_w) \tan \alpha + (q_1 - q_2) \Lambda_2 \right] \\
k_{\varphi\varepsilon}^c &= 0, & k_{\varphi\varepsilon}^l &= m E_w A_w (R_c + R_w) \left[ q_1 \tan \alpha + q_3 \sin \alpha + (-q_1/\tan \alpha + q_3 \sin \alpha) \Lambda_1 \right] \\
k_{\varphi\varphi}^c &= \mu_c J_c, & k_{\varphi\varphi}^l &= m E_w A_w (R_c + R_w) \left[ \left( q_1 + \frac{q_4}{\cos^3 \alpha} \right) (R_c + R_w) + (q_1/\tan \alpha - q_3 \sin \alpha) \Lambda_2 \right]
\end{aligned} \quad (13)$$

Also, the radial deformation $\rho$ of the helix centreline of the external wires in Eq.(6) can be rewritten as

$$\rho = A \varepsilon + B \varphi \quad (14)$$

with the coefficients $A$ and $B$ taking the form

$$\begin{aligned}
A &= \frac{1}{q_6} \left[ q_5 - \frac{\omega_3 (q_{12} + q_{13})}{64 E_c E_w} \right] \\
B &= \frac{1}{q_6} \left[ q_7 - \frac{\omega_4 (q_{14} + q_{15})}{64 E_c E_w} \right]
\end{aligned} \quad (15)$$

In the equations (13)-(15), the parameters $q_i$ can be explicitly related to geometrical and mechanical features of the strand as follows

$$\begin{aligned}
q_1 &= \cos^2 \alpha \sin \alpha, & q_2 &= q_1 \frac{R_w^2 \cos \alpha (1 + \nu_w \sin^2 \alpha)}{4 (R_c + R_w)^2 (1 + \nu_w)} \\
q_3 &= -q_1 \frac{R_w^2 \cos \alpha \left[ 1 + \nu_w (1 + \sin^2 \alpha) \right]}{4 (R_c + R_w)^2 (1 + \nu_w)} \\
q_4 &= q_1 \frac{R_w^2 \cos \alpha \left[ \sin^6 \alpha + (1 + \nu_w) \cos^2 \alpha (1 + \sin^2 \alpha)^2 \right]}{4 (R_c + R_w)^2 (1 + \nu_w)} \\
q_5 &= R_c \nu_c + R_w \nu_w \sin^2 \alpha, & q_6 &= R_c + R_w \nu_w (1 + \cos^2 \alpha) \\
q_7 &= R_w (R_c + R_w) \nu_w \cos \alpha \sin \alpha
\end{aligned} \quad (16)$$

and

$$q_8 = 2E_c\left(1-v_w^2\right)\log\left(\frac{1}{8(1-v_w)}+\sqrt{\frac{\omega_5}{|\omega_1|}}\right), \quad q_9 = 2E_w\left(1-v_c^2\right)\log\left(\omega_s - \frac{1}{2}+\frac{m}{2}-\sqrt{\frac{\omega_4}{|\omega_1|}}-\log 2\right)$$

$$q_{10} = 2E_c\left(1-v_w^2\right)\log\left(\frac{1}{8(1-v_w)}+\sqrt{\frac{\omega_5}{|\omega_2|}}\right), \quad q_{11} = 2E_w\left(1-v_c^2\right)\log\left(\omega_s - \frac{1}{2}+\frac{m}{2}-\sqrt{\frac{\omega_6}{|\omega_2|}}-\log 2\right)$$

$$q_{12} = 2E_c\left(v_w^2-1\right)\log\left(\frac{1}{8(v_w-1)}+\sqrt{\frac{\omega_7}{\varepsilon|\omega_3|}}\right), \quad q_{13} = 2E_w\left(v_c^2-1\right)\log\left(\frac{1}{2}-\omega_s-\frac{m}{2}+\sqrt{\frac{\omega_8}{\varepsilon|\omega_3|}}-\log 2\right) \quad (17)$$

$$q_{14} = E_c\left(v_w^2-1\right)\log\left(\frac{1}{8(v_w-1)}+\sqrt{\frac{\omega_7}{2\varphi|\omega_4|}}\right), \quad q_{15} = E_w\left(v_c^2-1\right)\log\left(\frac{1}{2}-\omega_s-\frac{m}{2}+\log 2+\sqrt{\frac{\omega_8}{2\varphi|\omega_4|}}\right)$$

in which

$$\Lambda_1 = \left(q_5 - \frac{q_8+q_9}{E_c}\omega_1\right)q_6^{-1}, \quad \Lambda_2 = \left(-q_7+\frac{q_{10}+q_{11}}{E_c}\omega_2\right)q_6^{-1} \quad (18)$$

and

$$\omega_1 = \frac{q_1 R_w^2}{R_c + R_w}\sin\alpha\left\{\left[1 - \frac{R_w^2 \cos^4\alpha}{4(R_c+R_w)^2(1+\nu_w)} + \frac{R_w^2 \cos^2\alpha \sin^2\alpha}{4(R_c+R_w)^2}\right] + \right.$$

$$\left. - \frac{\cos\alpha\left(\frac{2R_c\nu_c}{\tan\alpha} + R_w\nu_w\sin 2\alpha\right)}{2R_c + 2R_w(1+\nu_w\cos^2\alpha)}\left[1 + \frac{R_w^2 \cos^4\alpha \sin^2\alpha}{4(R_c+R_w)^2(1+\nu_w)} - \frac{R_w^2 \sin^4\alpha}{4(R_c+R_w)^2}\right]\right\}$$

$$\omega_2 = q_1 R_w^2 \cos\alpha\left\{\left[1 - \frac{R_w^2 \sin^2\alpha}{4(R_c+R_w)^2(1+\nu_w)} - \frac{R_w^2 \sin^2\alpha(1+\sin^2\alpha)}{4(R_c+R_w)^2}\right] + \right.$$

$$\left. - \frac{2R_w\nu_w \cos^2\alpha}{2R_c + 2R_w(1+\nu_w\cos^2\alpha)}\left[1 + \frac{R_w^2 \cos^2\alpha \sin^2\alpha}{4(R_c+R_w)^2(1+\nu_w)} - \frac{R_w^2 \sin^4\alpha}{4(R_c+R_w)^2}\right]\right\}$$

$$\omega_3 = \frac{8q_1 E_w \cos^2\alpha}{q_6(R_c+R_w)}\left\{-8R_c\nu_c/\sin\alpha + 8\tan\alpha(R_c+R_w)/\cos\alpha + \right.$$

$$\left. - \frac{R_w^2\left[R_c(1+\nu_c) + R_w(1+\nu_w)\right]\left[-\nu_w + (2+\nu_w)\cos 2\alpha\right]\sin\alpha}{(R_c+R_w)^2(1+\nu_w)}\right\}$$

$$\omega_4 = \frac{q_1 E_w \cos\alpha}{q_6(R_c+R_w)^2(1+\nu_w)}\left\{\left(32R_c^3 + 86R_cR_w^2 + 96R_c^2R_w\right)(1+\nu_w) + 3R_cR_w^2\nu_w + \right.$$

$$+ R_w^3\left[22 + 23\nu_w - \nu_w^2 + \nu_w(2+\nu_w)\cos 4\alpha\right] +$$

$$\left. + (R_c+R_w)\left[4R_w^2(3+2\nu_w)\cos 2\alpha - R_w^2(2+\nu_w)\cos 4\alpha\right]\right\}$$

$$\omega_5 = -\frac{E_c R_w(2R_c + R_w\sin^2\alpha)}{R_c\left[E_w(1-\nu_c^2) + E_c(1-\nu_w^2)\right]}, \quad \omega_6 = -\frac{E_c R_c(R_c + R_w\sin^2\alpha)}{R_c\left[E_w(1-\nu_c^2) + E_c(1-\nu_w^2)\right]} \quad (19)$$

$$\omega_7 = \frac{64 E_c E_w R_w(R_c + R_w\sin^2\alpha)}{R_c\left[E_w(1-\nu_c^2) + E_c(1-\nu_w^2)\right]}, \quad \omega_8 = \frac{64 E_c E_w(R_c + R_w\sin^2\alpha)}{E_w(1-\nu_c^2) + E_c(1-\nu_w^2)}$$

$$\omega_s = \sum_{j=1}^{\frac{m}{2}-1}\left\{\cos\left(\frac{2\pi j}{m}\right)\log\left[\tan\left(\frac{\pi j}{m}\right)\right] - \frac{\pi(1-2\nu_c)\sin(2\pi j/m)}{4(1-\nu_c)}\right\}$$

The constitutive matrix in Eq. (12) provides a non-symmetrical formal expression of the out-of-diagonal components, i.e. $k_{\varepsilon\varphi}^1 \neq k_{\varphi\varepsilon}^1$. This result is implicit for the vast majority of the approaches based on Costello's model as well as on other similar models [19, 41, 25], and is related to the constitutive aspect that links the radial stress to the in-plane shear displacement, although more recent works show the possibility of extending the Argatov-based approach by directly deriving a symmetric stiffness matrix [27].

The constitutive equations (12) can easily be generalised to strands with more complex hierarchical arrangements, such as multi-layer and multi-core strands illustrated in Figure 3 [38]. In fact, the particular structure of the stiffness matrix in Eq. (12) can be related to the case of a strand made of $n_1 \in \mathbb{N}$ layers, each of which constituted by $m_1 \in \mathbb{N}$ wires, by updating the stiffness coefficients as follows:

$$k_{\varepsilon\varepsilon}^l = \sum_{i=1}^{n_l} k_{\varepsilon\varepsilon}^i(m_i), \quad k_{\varepsilon\varphi}^l = \sum_{i=1}^{n_l} k_{\varepsilon\varphi}^i(m_i), \quad k_{\varphi\varepsilon}^l = \sum_{i=1}^{n_l} k_{\varphi\varepsilon}^i(m_i), \quad k_{\varphi\varphi}^l = \sum_{i=1}^{n_l} k_{\varphi\varphi}^i(m_i) \qquad (20)$$

where $i$ refers to the $i$-th layer and, accordingly, Eqs.(13)-(19) change depending on the corresponding number of wires $m_i \in \{1,...,m_l\}$ characterizing the $i$-th layer. Thus, by virtue of Eqs. (12) and (20), the general form of the constitutive relation for a strand with prescribed hierarchical microstructure can be written as

$$\boldsymbol{f} = \text{sym}[\mathbb{K}]\boldsymbol{d} \quad \Rightarrow \quad \begin{bmatrix} F \\ M_t \end{bmatrix} = sym \begin{bmatrix} \left(k_{\varepsilon\varepsilon}^c + \sum_{i=1}^{n_l} k_{\varepsilon\varepsilon}^i(m_i)\right) & \left(k_{\varepsilon\varphi}^c + \sum_{i=1}^{n_l} k_{\varepsilon\varphi}^i(m_i)\right) \\ \left(k_{\varphi\varepsilon}^c + \sum_{i=1}^{n_l} k_{\varphi\varepsilon}^i(m_i)\right) & \left(k_{\varphi\varphi}^c + \sum_{i=1}^{n_l} k_{\varphi\varphi}^i(m_i)\right) \end{bmatrix} \times \begin{bmatrix} \varepsilon \\ \varphi \end{bmatrix} \quad (21)$$

in which sym $[\mathbb{K}]$ is the actual strand stiffness matrix, $\boldsymbol{f} = [F, M_t]^T$ is the vector which collects the forces and $\boldsymbol{d} = [\varepsilon, \varphi]^T$ represents the vector of the overall strand deformation. Moreover, provided that $\det(\text{sym}\,\mathbb{K}) \neq 0$ (the physical meaning of the elastic strain energy actually leads to $\frac{1}{2}\boldsymbol{d}^T (\text{sym}\,\mathbb{K})\boldsymbol{d} > 0, \forall \boldsymbol{d} \neq 0 \Rightarrow \det(\text{sym}\,\mathbb{K}) > 0$), it is also possible to write inverse form of (21) as

$$\boldsymbol{d} = \mathbb{D}\,\boldsymbol{f}, \qquad \mathbb{D} \equiv [\text{sym}\,\mathbb{K}]^{-1} \qquad (22)$$

where $\mathbb{D}$ represents the compliance strand matrix. Both the constitutive Eqs. (21) and (22) are decisive in practical problems because they allow to treat cases where force-prescribed as well as displacement-prescribed boundary conditions need to be modelled. In fact, the lengthening-twisting coupling effect, due to the presence of nonzero out-of-diagonal coefficients in the stiffness strand matrix (21), inhibits the possibility of observing elongation and torsion as separate mechanisms in a strand with helical microstructure. Therefore, when axial forces are prescribed at the strand extremities, the deformation is generally characterized by elongation accompanied by twisting and, conversely, when the stretching is assigned in a test, the strand is stressed by both tensile forces and torsion.

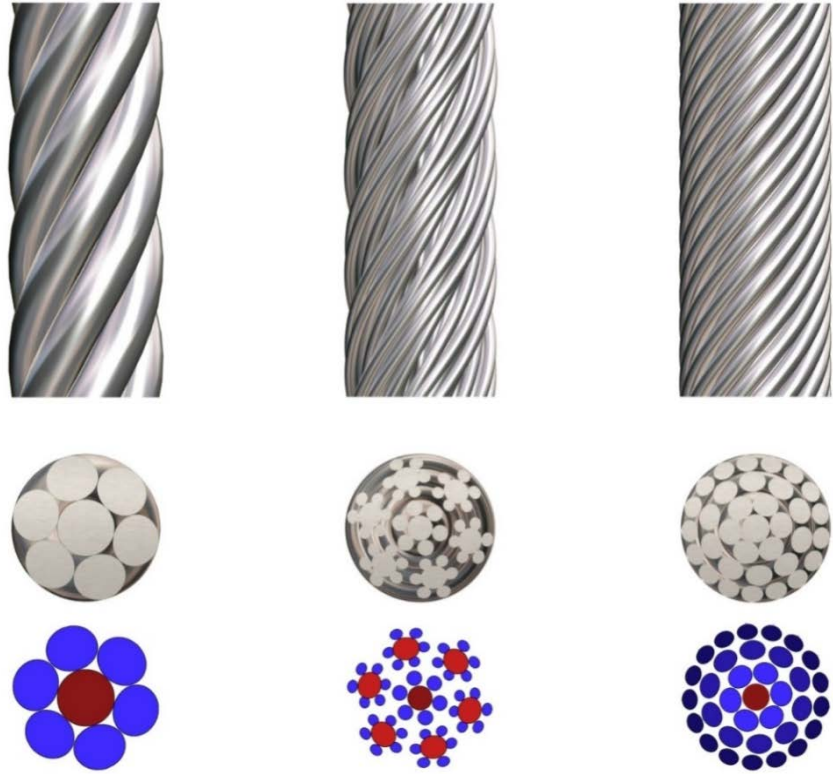

**Figure 3:** Representation of different hierarchical arrangements of wires and cores in a strand: simple straight strand (*left*), multi-core strand (*middle*) and multi-layer strand (*right*).

Moreover, the overall coupling between axial force and torsion in the strand is the effect of an analogous coupling at the wire level, where the additional bending regime participates to the kinematics. At this *local* microstructure scale level, the key geometrical parameters that play the main role in the deformation of the wire are the twisting angle $\alpha$, the ratio $R_w/R_c$ between wire and core radii and the helix pitch

$$p_h = 2\pi(R_c + R_w)\tan\alpha \qquad (23)$$

illustrated in Figure 2. As in the case of the strand, boundary conditions applied in terms of prescribed forces or displacements strongly influence the mechanical response of the wire, also producing some counterintuitive trends in the deformation when axial wire elongation $\varepsilon_w$, helix curvature variations $\Delta\kappa'_w$ and wire torsion variation $\Delta\tau_w$ are plotted against helix angle $\alpha$. These results are illustrated in Figure 4, where – under the hypotheses of linearly isotropic materials, small deformations and zero transversal contraction of the core ($\nu_c = 0$) – two complementary limit cases are considered. In both, the same tensile axial forces are applied. In particular, in the first case the twist at the ends is locked ($\varphi = 0, M_t \neq 0$), while in the second case the ends are free ($\varphi \neq 0, M_t = 0$). The plots show the three wire kinematical parameters, $\varepsilon_w$, $\Delta\kappa'_w$ and $\Delta\tau_w$, evolving with the initial helix angle $\alpha \in (0, \pi/2)$. Although geometrical compatibility requires to confine the helix angle variation within $\{\alpha \in [\alpha_{\min}, \pi/2[, \ \alpha_{\min} \equiv \arctan[R_w/(\pi R_c + \pi R_w)] > 0\}$, it is interesting to observe that the maximum of wire elongation $\varepsilon_w$ corresponds to the overall strand stretching $\varepsilon$, while the wire bending curvature $\Delta\kappa'_w$ exhibits non-monotonic variations and the wire twist $\Delta\tau_w$ shows a nonlinear and non-monotonic trend which is also accompanied by an unexpected change in sign in the proximity of $\alpha \approx \pi/4$, to accommodate the geometrical congruency due to the mutual interaction between local torsion and bending curvatures.

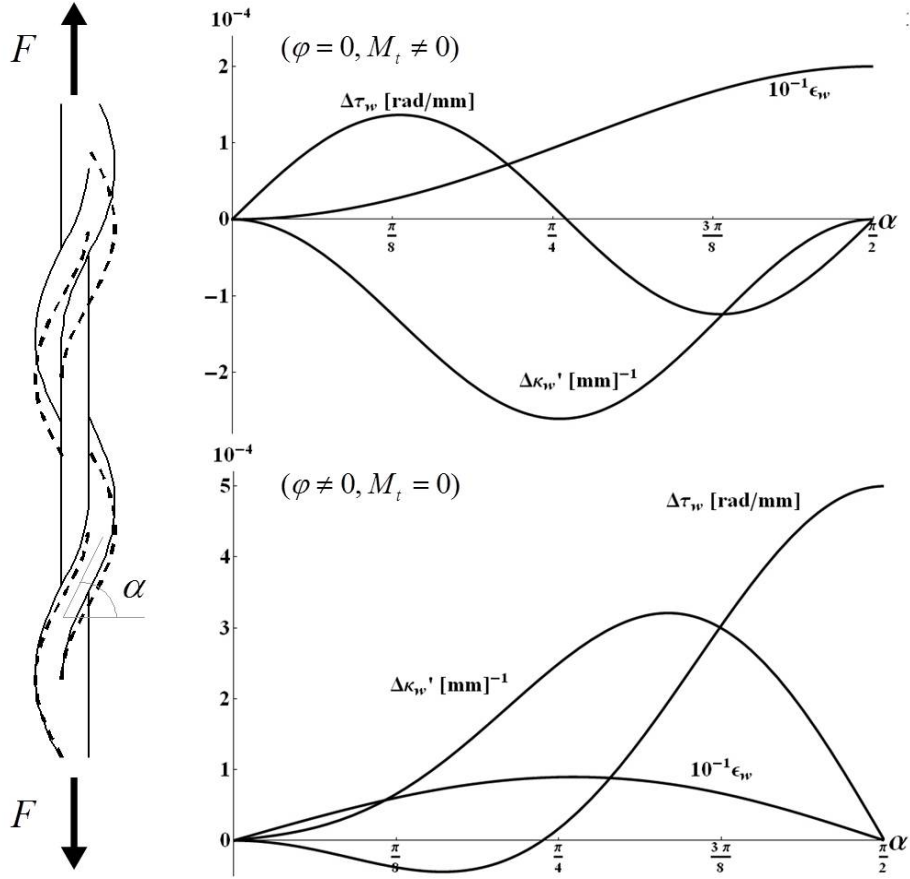

**Figure 4:** Deformations of a single wire in a strand at overall prescribed force $F = 40\,\text{kN}$ versus varying helical angle, for two complementary boundary conditions imposed at strand ends. (Top) Torsion per unit length looked at the strend extremities ($\varphi = 0$). (Bottom) Vanishing overall torque imposed at the strand ends ($M_t = 0$). Strand parameters used in the calculations: $R_c = 1.97$ mm, $R_w = 1.865$ mm, $E_c = 197.9$ GPa, $v_c = 0$, $v_w = 0.3$.

## 3. Derivation of the equivalent monoclinic/trigonal cylindrical strand

If the strand is seen as a whole, the coupling between axial force and torsion can be captured by assuming that the equivalent homogenized cylinder comprising the underlying strand structure is an anisotropic monoclinic/trigonal material. The mathematically rigorous identification of the mechanical behaviour of a strand structure with an equivalent continuum cylinder whose material exhibits monoclinic/trigonal symmetries allows to exploit closed-form analytical solutions from Functionally Graded Material cylinders made of *n* hollow monoclinic layers and a central isotropic core developed by Cutolo et al. [43]. Thus, a direct correspondence can be established between the constitutive equations (21) of the strand structure, here relative to the case of a single straight strand, and the elastic moduli of the continuum trigonal cylinder. Therefore, a homogenized model that macroscopically behaves as the actual strand is constructed, governed by the same microstructural material and geometrical parameters, i.e. elastic moduli, number and *architecture* of wires and cores within the strand.

### 3.1. Monoclinic/trigonal material assumptions

According to Lord Kelvin's definition [44], a structure is defined *chiral* if it cannot be superposed on its mirror image, i.e the structure presents a handedness [45]. Helical spirals constitute an example of *chirality*. In the framework of Continuum Mechanics, media exhibiting underlying chiral microstructures macroscopically show a *monoclinic* behaviour and the classification of this peculiar type of anisotropy depends on the number and orientation of the vectors normal to the planes of material symmetry, in full analogy with the crystallographic method encountered in Group Theory. In particular, the monoclinic crystal system has exactly one plane of reflective symmetry and the elastic symmetry retains, in its canonical material reference system, cross-elastic constants relating normal stresses (strains) to shear strains (stresses). In the well-known Voigt notation, the $(6 \times 6)$ elasticity tensor $\mathsf{C}$ with respect to a cylindrical reference system $\{r, \vartheta, x_3\}$ where $x_3$ is the axis of the cylinder, assumes the form

$$\boldsymbol{T} = \mathsf{C} : \boldsymbol{E} \quad \Rightarrow \quad \begin{bmatrix} \sigma_{rr} \\ \sigma_{\theta\theta} \\ \sigma_{33} \\ \sigma_{3\theta} \\ \sigma_{3r} \\ \sigma_{\theta r} \end{bmatrix} = \begin{bmatrix} c_{11} & c_{12} & c_{13} & c_{14} & 0 & 0 \\ c_{12} & c_{22} & c_{23} & c_{24} & 0 & 0 \\ c_{13} & c_{23} & c_{33} & c_{34} & 0 & 0 \\ c_{14} & c_{24} & c_{34} & c_{44} & 0 & 0 \\ 0 & 0 & 0 & 0 & c_{55} & c_{56} \\ 0 & 0 & 0 & 0 & c_{56} & c_{66} \end{bmatrix} \times \begin{bmatrix} \varepsilon_{rr} \\ \varepsilon_{\theta\theta} \\ \varepsilon_{33} \\ 2\varepsilon_{3\theta} \\ 2\varepsilon_{3r} \\ 2\varepsilon_{\theta r} \end{bmatrix} \quad (24)$$

in which $[\boldsymbol{T}]_{ij} = \sigma_{ij}$ and $[\boldsymbol{E}]_{ij} = \varepsilon_{ij}$ are the stress and the strain tensors, respectively. Contrary to other more classical types of elastic anisotropy (i.e. cubic, transverse isotropy and orthotropy) the characteristic feature of the monoclinic/trigonal materials considered here is the symmetry-breaking character of the cross-elastic constants $c_{14}$, $c_{24}$ and $c_{34}$, $c_{34}$. These coefficients are directly responsible for the coupling between the normal stress $\sigma_{33}$ and the shear strain $\varepsilon_{3\theta}$ generated by the presence of torsion. Therefore, if an ideal cylinder comprising the strand microstructure is taken as an equivalent continuum model, some key geometrical properties of the actual strand architecture must be preserved. As a consequence, the overall isotropy in the strand cross section due to the core and wire arrangement is captured by means of a constraint among the corresponding isotropic elastic coefficients $c_{11}$, $c_{22}$ and $c_{12}$. The above-mentioned symmetry-breaking cross-elastic coefficient $c_{34}$ is instead the natural candidate to represent the coupling between elongation and twisting that characterizes the mechanical response of the strand. This means that, as the wire helix angle $\alpha$ tends to $\pi/2$, the elasticity tensor of the equivalent continuum cylinder in Eq. (24) is expected to reduce to that of hexagonal or transversely isotropic materials [46].

### 3.2. Axial-torsional effects in an elastic monoclinic/trigonal cylinder

We recall the governing equations for the displacement field of an elastic hollow cylinder made of a monoclinic/trigonal material. The analytical derivation of the displacement solutions, extensively developed in Cutolo et al. [43] also on account of the analytical formulation provided by Fraldi et al. [47], is reported in Appendix A. By considering elasto-static problems in the cylindrical reference system $\{r, \vartheta, x_3\}$ for which the stress components can be assumed independent of the $x_3$ variable, the components of the displacement vector $\boldsymbol{u}$, with $u_i$, $i \equiv \{r, \vartheta, x_3\}$, assume the form [43]:

$$u_r(r) = C_1\left(r^{h_0} + r^{-h_0}\right) + C_2\left(r^{h_0} - r^{-h_0}\right) + h_1\, r\,\varepsilon + h_2\,\varphi\, r^2,$$
$$u_\vartheta(r, x_3) = r\, x_3\, \varphi + h_4\, C_4 - h_3\, C_3 / r, \qquad (25)$$
$$u_3(r, x_3) = x_3\, \varepsilon + h_4\, C_3 / r + h_5\, C_4 \log(r),$$

where $C_1$, $C_2$, $C_3$ and $C_4$ are integration constants, $\varepsilon$ and $\varphi$ are respectively the axial strand deformation and the torsional angle per unit length of the cylinder, and the coefficients $h_j$ read as [43]

$$h_0 = \sqrt{\frac{c_{22}}{c_{11}}}, \quad h_1 = \frac{c_{23} - c_{13}}{c_{11} - c_{22}}, \quad h_2 = \frac{c_{24} - 2c_{14}}{4c_{11} - c_{22}},$$
$$h_3 = \frac{c_{55}}{2\left(c_{56}^2 - c_{55}\, c_{66}\right)}, \quad h_4 = \frac{c_{56}}{c_{56}^2 - c_{55}\, c_{66}}, \quad h_5 = \frac{c_{66}}{c_{56}^2 - c_{55}\, c_{66}} \qquad (26)$$

Starting from Eq. (25), the strains $\varepsilon_{ij}$ and the stresses $\sigma_{ij}$ are easily derived by means of the compatibility equations $\boldsymbol{E} = sym(\mathbf{u} \otimes \nabla)$ written in the cylindrical coordinate system at hand (see the Appendix A) and constitutive laws in Eq. (24).

### *3.3. Axial force and torque*

In the present problem, the strand can be seen as 2-FGMC with circular cross-section subjected to the combination of axial force and torque, so that the equivalent homogenized elastic constitutive behaviour can be derived in a completely analytical way. In particular, an elastic heterogeneous cylinder made of a cylindrical core of isotropic material surrounded by a layer of trigonal material is here considered. The core of the FGMC reproduces the strand core, while the outer monoclinic/trigonal layer mimics the overall response of the wounded external wires of the strand. The double-layer cylinder then represents an equivalent homogenized model for the simple straight strand. The equivalent model is a homogenized solid cylinder made up of an isotropic cylindrical core of radius $R_c$ embedded in an ideal cladding constituted by a trigonal hollow cylinder of inner radius $R_c$ and outer radius $R_t = R_c + 2R_w$. The mechanical equivalence between the double-layer cylinder and the simple straight strand is sought through the homogenized material elastic constants $c_{ij}^c$ and $c_{ij}^l$, where $c$ and $l$ are referred to the core and the layer of wires (the overall hollow cylindrical phase), respectively. Then, from Eqs. (24), the mechanical behaviour of the core is defined through the isotropic elastic moduli

$$c_{11}^c = c_{22}^c = c_{33}^c = \lambda_c + 2\mu_c$$
$$c_{12}^c = c_{13}^c = c_{23}^c = \lambda_c \qquad (27)$$
$$c_{44}^c = c_{55}^c = c_{66}^c = \mu_c$$

where $\lambda_c$ and $\mu_c$ are the Lamé constants of the isotropic core material, all the other elastic values $c_{ij}^c$ being vanishing due to isotropy [48]. With reference to the external phase, since the cross sectional plane of the strand represents a plane of isotropy for the strand mechanical behaviour [32, 42, 33, 46], the general equations (24) are representative of the equivalent mechanical model of the wrapping wires if the following relations between the elastic constants $c_{ij}$ are introduced [49]

$$\begin{aligned} c_{11}^l &= c_{22}^l \\ c_{66}^l &= \frac{c_{11}^l - c_{12}^l}{2} \end{aligned} \qquad (28)$$

and

$$c_{13}^l = c_{23}^l, \quad c_{14}^l = c_{24}^l, \quad c_{44}^l = c_{55}^l, \quad c_{56}^l = 0 \qquad (29)$$

The cylindrical reference system is placed on the centre of the mass, the axis of the cylinder is taken to lay on the $x_3$ direction, coincident with the longitudinal axis of the cylinder. Self-balancing forces $\overline{F}$ and self-balancing torque couples $\overline{M}_t$ are applied at the bases of the cylinder, so that equilibrium considerations lead to write

$$\begin{aligned} \overline{F} &= \overline{F}^c + \overline{F}^l \\ \overline{M}_t &= \overline{M}_t^c + \overline{M}_t^l \end{aligned} \qquad (30)$$

The displacement field in the isotropic inner layer is [43]

$$\begin{aligned} u_r^c(r) &= C_0\, r \\ u_\theta^c(r, x_3) &= r\, x_3\, \varphi^c \\ u_3^c(r, x_3) &= x_3\, \varepsilon^c \end{aligned} \qquad (31)$$

for $r \in [0, R_c]$, where $C_0$ is an integration constant (see Appendix A). Therefore, Eq. (31) together with the displacement components $u_i^l$, $i = \{r, \vartheta, x_3\}$ having the expression given in the Eqs. (25), defined for $r \in [R_c, R_t]$ and limited to the monoclinic/trigonal layer l by means of the restrictions in Eqs. (28)-(29), constitute the analytical solution of the strand equivalent model. Then, by starting from Eqs. (25) and (31), the strain and the stress fields are finally derived, the coefficients $C_i$, $i = 0,...,4$, the elongations $\varepsilon^c$ and $\varepsilon^l$ as well as the twisting angles per unit length $\varphi^c$ and $\varphi^l$, representing the 9 integration constants of the elastic problem to be found through suitable interface and boundary conditions. More precisely, we have 6 interface conditions at $r = R_c$, defined by the continuity of displacement and Cauchy stress vector at the corresponding surfaces:

$$\begin{cases} u_r^c(r=R_c) = u_r^l(r=R_c) \\ u_\vartheta^c(r=R_c) = u_\vartheta^l(r=R_c) \\ u_3^c(r=R_c) = u_3^l(r=R_c) \\ \sigma_{rr}^c(r=R_c) = \sigma_{rr}^l(r=R_c) \\ \sigma_{r3}^c(r=R_c) = \sigma_{r3}^l(r=R_c) \\ \sigma_{r\vartheta}^c(r=R_c) = \sigma_{r\vartheta}^l(r=R_c) \end{cases} \qquad (32)$$

Eqs. (32)$_{2, 3, 5, 6}$ imply that $\varepsilon^c = \varepsilon^l = \varepsilon$ and $\varphi^c = \varphi^l = \varphi$ as well as $C_3 = C_4 = 0$ and so the system in Eq. (32) reduces to the sole equilibrium equation at the interface between core and external layer, that is

$$\sigma_{rr}^c(r, x_3) = \sigma_{rr}^l(r, x_3), \quad r = R_c, \qquad (33)$$

while the continuity of the displacements at the core-layer cylindrical interface requires that

$$u_r^c(r) = u_r^l(r), \quad r = R_c \qquad (34)$$

to which the traction-free condition at the outermost cylindrical surface has to be added, i.e.:

$$\sigma_{rr}^l(r=R_t) = 0 \qquad (35)$$

Eqs.(33)-(35) can be rearranged in matrix form in a way to calculate the remaining unknown constants $C_0, C_1$ and $C_2$ as functions of the axial strain $\varepsilon$ and of the twisting angle per unit length $\varphi$ (their final expression is reported in Appendix A). Therefore, the stress field in the homogenized double-layer cylinder is now completely defined, depending upon geometric and material properties of the core and of the wires belonging to the external layer.

By starting from prescribed deformations, e.g. $\varepsilon$ and $\varphi$, one can write:

$$\begin{bmatrix} \overline{F} \\ \overline{M}_t \end{bmatrix} = \begin{bmatrix} \overline{F}^c + \overline{F}^l \\ \overline{M}_t^c + \overline{M}_t^l \end{bmatrix} = \begin{bmatrix} \overline{k}_{\varepsilon\varepsilon}^c + \overline{k}_{\varepsilon\varepsilon}^l & \overline{k}_{\varepsilon\varphi}^c + \overline{k}_{\varepsilon\varphi}^l \\ \overline{k}_{\varphi\varepsilon}^c + \overline{k}_{\varphi\varepsilon}^l & \overline{k}_{\varphi\varphi}^c + \overline{k}_{\varphi\varphi}^l \end{bmatrix} \begin{bmatrix} \varepsilon \\ \varphi \end{bmatrix}, \qquad (36)$$

In this system, the axial force $\overline{F}$ and the twisting moment $\overline{M}_t$ follow from Eq. (30), recalling that

$$\begin{aligned} \overline{F}^c &= 2\pi \int_0^{R_c} \sigma_{33}(x_3=0) \, r \, dr, & \overline{F}^l &= 2\pi \int_{R_c}^{R_t} \sigma_{33}(x_3=0) \, r \, dr, \\ \overline{M}_t^c &= 2\pi \int_0^{R_c} \sigma_{3\theta}(x_3=0) \, r^2 dr, & \overline{M}_t^l &= 2\pi \int_{R_c}^{R_t} \sigma_{3\theta}(x_3=0) \, r^2 dr. \end{aligned} \qquad (37)$$

while the coefficients $\overline{k}_{ij}^c$ and $\overline{k}_{ij}^l$ can be expressed as:

$$\bar{k}_{\varepsilon\varepsilon}^c = E_c A_c, \quad \bar{k}_{\varepsilon\varepsilon}^l = \frac{2\pi R_w r_h}{C_a} \left\{ \begin{array}{l} R_w r_h \left[ 2c_{33}^l \left( c_{11}^{l\,2} - c_{12}^{l\,2} \right) - c_{13}^l \left( c_{11}^l - c_{12}^l \right) \left( 4c_{13}^l + \dfrac{R_c^2}{R_w r_h} \lambda_c \right) \right] + \\ + \left[ 2c_{33}^l \left( c_{11}^l R_c^2 + 2 R_w r_h \left( c_{11}^{l\,2} + c_{12}^{l\,2} \right) \right) - 2c_{13}^{l\,2} R_t^2 \right] (\lambda_c + \mu_c) \end{array} \right\}$$

$$\bar{k}_{\varepsilon\varphi}^c = 0, \quad \bar{k}_{\varepsilon\varphi}^l = \frac{4\pi R_w \left( 3 r_h^2 + R_w^2 \right)}{3 C_a} \left\{ \begin{array}{l} R_w r_h \left[ c_{34}^l \left( c_{11}^{l\,2} - c_{12}^{l\,2} \right) - 2 c_{13}^l c_{14}^l \left( c_{11}^l - c_{12}^l \right) \right] + \\ + \left[ c_{34}^l \left( c_{11}^l R_c^2 + 2 R_w r_h \left( c_{11}^l + c_{12}^l \right) \right) - c_{13}^l c_{14}^l R_t^2 \right] (\lambda_c + \mu_c) \end{array} \right\}$$

$$\bar{k}_{\varphi\varepsilon}^c = 0, \quad \bar{k}_{\varphi\varepsilon}^l = \bar{k}_{\varepsilon\varphi}^l - \frac{2\pi \left( c_{11}^l - c_{12}^l \right) c_{14}^l R_c^2 R_w \left( 3 r_h^2 + R_w^2 \right) \lambda_c}{3 C_a}$$

$$\bar{k}_{\varphi\varphi}^c = \mu_c J_c, \quad \bar{k}_{\varphi\varphi}^l = \frac{16 \pi R_w}{36 c_{11}^l C_a} \left\{ \begin{array}{l} 9 c_{11}^l c_{44}^l R_w r_h^2 \left( r_h^2 + R_w^2 \right) \left( c_{11}^{l\,2} - c_{12}^{l\,2} \right) + c_{14}^{l\,2} R_w \left[ c_{12}^{l\,2} R_w^2 \left( 3 r_h^2 - R_w^2 \right) + \\ + 2 c_{11}^l c_{12}^l \left( 3 r_h^2 + R_w^2 \right)^2 - c_{11}^{l\,2} \left( 6 r_h \left( 12 R_c^3 + 17 R_w^3 \right) + \\ - 2 \left( 27 R_c^4 + 34 R_w^3 \right) + 123 R_c^2 R_w^2 \right) \right] + \left[ 9 c_{11}^l c_{44}^l \left( r_h^2 + R_w^2 \right) \left( 2 c_{12}^l R_w r_h^2 + \\ + c_{11}^l r_h \left( r_h^2 + R_w^2 \right) \right) - 2 c_{12}^l c_{14}^{l\,2} R_w^3 \left( 3 r_h^2 - R_w^2 \right) + \\ - 2 c_{11}^l c_{14}^{l\,2} \left( 9 R_c^4 \left( R_c + 7 R_w \right) + 12 R_c^2 R_w^2 \left( 15 R_c + 22 R_w \right) \right] + \\ + 68 R_w^4 \left( 3 R_c + R_w \right) \right) \right] (\lambda_c + \mu_c) \end{array} \right\}$$

(38)

where $r_h = R_c + R_w$, and $C_a = \left( c_{11}^{l\,2} - c_{12}^{l\,2} \right) R_w r_h + \left[ 2 \left( c_{11}^l + c_{12}^l \right) R_w r_h + c_{11}^l R_c^2 \right] (\lambda_c + \mu_c)$.

It is worth to highlight that the term $\dfrac{\bar{k}_{\varepsilon\varphi}^l - \bar{k}_{\varphi\varepsilon}^l}{\bar{k}_{\varepsilon\varphi}^l + \bar{k}_{\varphi\varepsilon}^l}$ describes the asymmetry of the constitutive matrix (12), which is related to the axial-twisting coupling. In this analytical model of a monoclinic/trigonal double layer cylinder, it directly depends on the stiffness term $c_{14}^l$, as shown below. In other words, among all the stiffness constants in the monoclinic constitutive relation (24) that accounts for the axial-torsional coupling in the system, i.e. $c_{14}^l$, $c_{24}^l$, $c_{34}^l$, the elastic coefficient $c_{14}^l$ is responsible for the fact that the tensile force-torsion and torque-elongation relations are not symmetric.

Using Eqs.(6) and (14), the radial deformation of the cylinder can be expressed at $r = R_t$ by the formula

$$\bar{\rho} = 1 - \frac{R_{t1}}{R_t} = -\frac{u_r^w(R_t)}{R_t} = \bar{A} \varepsilon + \bar{B} \varphi \tag{39}$$

where the solution of Eq. (25) has been employed. The constants in Eq. (39) can be written as:

$$\overline{A} = \frac{1}{C_b} \left\{ \begin{array}{l} -2\left(c_{11}^l - c_{12}^l\right) R_w \left[ -8c_{13}^l r_h^3 + R_c^2 \left(R_c + 3R_t\right) \lambda_c \right] + \\ +8R_t^2 \left[ c_{11}^l R_c^2 \lambda_c + c_{13}^l R_w \left(R_c + r_h\right)\left(\lambda_c + \mu_c\right) \right] \end{array} \right\}$$

$$\overline{B} = \frac{1}{6c_{11}^l C_b} \left\{ \begin{array}{l} c_{14}^l R_w \left[ c_{11}^{l\,2} \left(48 r_h^4 - 8R_c R_t^2\right) + c_{12}^l \left(4c_{12}^l R_w^2 \left(2R_c R_t + 4r_h^2\right) - 16 c_{11}^l r_h^2 \left(4r_h^2 - R_c R_t\right)\right) + \\ -\left(8 c_{12}^l R_w^2 \left(2R_c R_t + 4r_h^2\right) - 4c_{11}^l r_h \left(8 r_h^3 - 2R_t \left(4r_h^2 - 2R_c^2\right)\right)\right)\left(\lambda_c + \mu_c\right) \right] \end{array} \right\}$$

(40)

Where $C_b = 4r_h^2 \left[ \left(c_{11}^{l\,2} - c_{12}^{l\,2}\right)\left(R_c^2 - R_t^2\right) - 2\left(c_{11}^l \left(R_c^2 + R_t^2\right) - c_{12}^l \left(R_c^2 - R_t^2\right)\right)\left(\lambda_c + \mu_c\right) \right]$.

### 3.4. Strand equivalent model

The approach to describe the strand mechanics proposed by Argatov [36] allows to evaluate force and torque on a strand of wires once axial elongation and twisting angle are prescribed, see Eq. (12). It is based on the definition of the material properties of the wire and the geometrical arrangement of the strand. The same isotropic material is here assumed for the core and the wires. In case of a simple straight strand, the geometrical parameters are thus: the radius of the core $R_c$, the radius of the helically arranged wires $R_w$, and the helix pitch angle $\alpha$. Then, the comparison between Eq. (12) and Eq. (37) leads to the identification of forces and twisting moments as follows

$$\begin{array}{ll} \overline{F}^c = F^c, & \overline{F}^l = F^l, \\ \overline{M}_t^c = M_t^c, & \overline{M}_t^l = M_t^l, \end{array} \quad (41)$$

Eqs.(41) provide the constitutive law for the equivalent cylinder in both the core and the external layer, thanks to which the elastic constants $c_{ij}$ can be expressed as a function of the wire mechanical and geometrical features. In particular, by virtue of Eq. (12), the conditions (41) depend directly on constitutive constants prescribed by Argatov model, see (13), and are verified in case of trigonal elastic stiffness terms defined as:

$$c_{33}^l = \frac{k_{\varepsilon\varepsilon}^l}{4\pi r_h R_w} + \frac{c_{13}^l}{C_c}\left[\frac{c_{11}^l - c_{12}^l}{2}\left(4 r_h R_w c_{13}^l + R_c^2 \lambda_c\right) + c_{13}^l R_t^2 \left(\lambda_c + \mu_c\right)\right]$$

$$c_{34}^l = \frac{3 k_{\varepsilon\varphi}^l}{4\pi R_w \left(R_t^2 + 2 r_h R_c\right)} + \frac{c_{13}^l c_{14}^l}{C_c}\left[2 r_h R_w \left(c_{11}^l - c_{12}^l\right) + R_t^2 \left(\lambda_c + \mu_c\right)\right]$$

$$c_{44}^l = \frac{k_{\varphi\varphi}^l}{4\pi r_h R_w \left(R_t^2 - 2 r_h R_w\right)} + \frac{c_{14}^{l\,2} R_w}{9 c_{11}^l r_h \left(R_t^2 - 2 r_h R_w\right) C_c}\left[\left(\left(c_{11}^l - c_{12}^l\right) + 2\left(\lambda_c + \mu_c\right)\right)\right. \quad (42)$$

$$\left. \left(R_w^2 \left(c_{11}^l + c_{12}^l\right)\left(3 r_h^2 - R_w^2\right) + 2 c_{11}^l \left(R_t^2 + 2 r_h R_c\right)^2\right) + 9 c_{11}^l R_c^2 \left(R_c^2 + 2 r_h^2 + R_c^2 / R_w\right)\left(\lambda_c + \mu_c\right)\right]$$

$$c_{14}^l = \frac{3 C_a \left(k_{\varepsilon\varphi}^l - k_{\varphi\varepsilon}^l\right)}{2\pi \left(c_{11}^l - c_{12}^l\right) R_c^2 R_w \left(3 r_h^2 + R_w^2\right) \lambda_c}$$

where $C_c = \left(c_{11}^{l\,2} - c_{12}^{l\,2}\right) R_w r_h - \left[2\left(c_{11}^l - c_{12}^l\right) R_w r_h - c_{11}^l R_t^2\right](\lambda_c + \mu_c)$. The expression of the radial deformation permits to identify the constants $c_{13}^l$ and $c_{14}^l$. In fact, by imposing the equivalence between Eq. (14) and Eq. (39), one observes that

$$\overline{A} = A, \quad \overline{B} = B, \tag{43}$$

These relations provide the constitutive law for the equivalent cylinder in the external layer

$$c_{13}^l = \frac{2\overline{A} r_h R_w \left[\left(c_{11}^{l\,2} - c_{12}^{l\,2}\right) + 2\left(c_{11}^l + c_{12}^l\right)(\lambda_c + \mu_c)\right] + c_{11}^l R_c^2 \left((\lambda_c + \mu_c) 2\overline{A} - \lambda_c\right)}{2\pi r_h R_w \left[c_{11}^l - c_{12}^l + 2(\lambda_c + \mu_c)\right]}$$

$$c_{14}^l = \frac{3\overline{B} r_h R_w \left[\left(c_{11}^{l\,2} - c_{12}^{l\,2}\right) + 2\left(c_{11}^l + c_{12}^l\right)(\lambda_c + \mu_c)\right] + c_{11}^l R_c^2 (\lambda_c + \mu_c) 3\overline{B}}{R_w \left(3 r_h^3 + R_w^2\right)\left[c_{11}^l - c_{12}^l + 2(\lambda_c + \mu_c)\right]}$$

(44)

It can be verified that the last of (42) corresponds to the second of (44). In order to find the remaining elastic constants of the monoclinic layer, we exploit the hypothesis of considering the homogenized behaviour of the external layer in the plane $r\text{-}\theta$ as isotropic. This assumption allows to apply a suitable homogenization technique for an isotropic hollow cylinder under external pressure (see Appendix B). As a result, the in-plane mechanical behaviour is defined so that

$$c_{11}^l = \gamma^n \frac{16 r_h R_w \mu_w \left[R_c^2 (\lambda_w + \mu_w) + R_c R_w (\lambda_w + 2\mu_w) + R_w^2 (\lambda_w + 2\mu_w)\right]}{\left(R_w^2 + r_h^2\right)\left[8\mu_w R_w r_h + R_c^2 (3\lambda_w + 4\mu_w)\right]},$$

$$c_{12}^l = \gamma^n \frac{4 r_h R_w R_t^2 \lambda_w \mu_w}{\left(R_w^2 + r_h^2\right)\left[8\mu_w R_w r_h + R_c^2 (3\lambda_w + 4\mu_w)\right]}, \quad c_{66}^l = \gamma^n \frac{2 r_h R_w}{R_w^2 + r_h^2} \mu_w,$$

(45)

where $\gamma = n R_w \left[4(R_c + R_w)\sin\alpha\right]^{-1}$ is the volume fraction given as the ratio between the bulk volume of the external wires and the volume of the equivalent hollow cylinder, and $n$ is a parameter to be fixed (we set $n = 2$ in all results below). It is also worth recalling that the constants in Eq. (45) follow from the conditions in Eq. (28).

## 4. Implementation of monoclinic/trigonal homogenized strand model in reinforced composites

One of the most common applications of wire rope strands – at different scale levels – is in the field of reinforced macro-, micro- and nano-composites characterized by short or long fibres with helical microstructure. The reinforcing fibres are designed to be embedded in a linear or nonlinear elastic isotropic matrix, with perfect bonding at the interfaces to guarantee the best mechanical performance in terms of overall stiffness. However, the possible monoclinic/trigonal feature of the reinforcing constituents is generally neglected in the common design procedures and thus in the calculations aimed to predict the structural response of the composites. The studies in the literature to determine the stiffness (and the strength) of composite materials are usually performed by taking into account the sole contribution of the fibres along their axis, somehow including the effects of possible anti-plane interfacial shear stresses, but without considering the influence of possible shear stresses emerging at the cylindrical reinforcement/matrix interfaces along the hoop direction. At least from the theoretical point of view, these other "spurious" shear stresses may occur when the microstructure of the reinforcing elements is characterized by helical/chiral structures as a consequence of the axial-torsional coupling that governs the monoclinic/trigonal elastic behaviour of the fibres. In these cases, the magnitude and spatial distribution of shear stresses transferred to the matrix can be very different from those predicted by assuming that the homogenized behaviour of the cords simply obeys isotropic or orthotropic elastic laws, with important potential effects on the qualitative and quantitative estimation of phenomena like detaching at the interfaces, onset of damaging and fatigue loads in a wide class of materials such as pneumatic tires, air springs, belt structures and nano-composites in service conditions. As previously mentioned, the twisted nature of the cords generates a significant axial-torsional coupling effect [50]. The main mechanical features of composites are strongly dependent upon the arrangement of the filaments in the cord. Therefore, it is essential to evaluate how extension-twisting behaviour affects the effective structural response of cord-rubber composite materials, especially when it is necessary to reduce the failure risk related to fatigue. For example, the presence of stresses generated at the cord-rubber interface, apart from its importance for the general understanding of the structural behaviour of rubber composites, is the main cause of composite failure, and its knowledge is of key importance in their design. In fact, at microscale, the interaction between the cord and rubber triggers failure phenomena and significant stiffness discrepancies of the constituents and the related strong stress gradients can produce stress concentrations at the material interfaces, promoting crack propagation. In rubber-like matrices, for instance, due to incompressible behaviour of the material, the matrix is strongly vulnerable to deviatoric components of stress [51] and hence shear stresses, induced by the presence of the reinforcing fibres with a monoclinic/trigonal behaviour, play a crucial role in the overall response of the composite. Furthermore, shear stresses transferred across the material interfaces are also relevant in other types of composites, at different scales.

To analyse in detail the influence of the above-mentioned stress fields and to show the importance of the proposed method, we consider a case study represented by a composite unit constituted by a single ply in which unidirectional reinforced trigonal/monoclinic structures are embedded. Results of FE analysis highlight how shear stresses due to monoclinic cord behaviour propagate to the surrounding matrix and increase with unexpected peaks close to the material interface. Since the distance between cords as well as their mutual orientation represent important design characteristics that largely influence the magnitude of such effects, as demonstrated for example in experimental tests in the literature [52], a sensitivity analysis is additionally provided in the simulations, to highlight multiplicative effects related to nonlinear superposition of shear stresses that could contribute to explain failure phenomena and help optimal design procedures for fiber-reinforced laminates [53].

### *4.1. Finite element model of the pad unit of a composite*

Although the present model applies to various different types of composites, to highlight the effectiveness and the possible relevance of the observations made on the influence of shear stress transfer at the (trigonal) fibre-matrix interface, a parametric model of reinforcing cords embedded in a rubber matrix –of interest for example in tire applications – has been considered and then implemented by means of an *ad hoc* FE model in Ansys® [54], using simple single strands. The adopted reinforcing fibres consist of 6 wires wrapped around a straight core wire with a lay angle $\alpha$ (Figure 2). The strand is modelled through an equivalent cylinder, as discussed in the previous sections. Once the overall properties of the equivalent cylinder are calculated by means of Eqs. (27) to (42), the homogenized linear elastic behaviour of the cord is implemented in a FE model, thereby avoiding the need to geometrically reconstruct the detailed strand microstructure and drastically reducing the related computational costs. The adoption of an equivalent monoclicic/trigonal cylinder provides the considerable advantage of using very few elements for reconstructing the strands, while preserving their peculiar behaviour – characterized by the coupling of torsion and elongation – without reproducing the large number of wires disposed in a multi scale arrangement (Multi Layered

Strand). This numerical system accounts for the intrinsic trigonality of the simple strand at a lower scale level, thereby projecting the global trigonal effect due to cord-rubber interaction in the composite at higher scales. A fully parametric procedure has been implemented, by assuming a perfect interface between the cords and the rubber. The axial-twisting coupling constants of the homogenized FE cylinder, in very good agreement with those furnished by the analytical model by Argatov for different helix angles, are reported in Figure 5.

The mechanical behaviour of the FE model strictly reproduces the response of the analytical model of the homogenized cylinder. FE simulations of monoclinic/trigonal cylinders embedded in a rubber matrix are then performed. The implemented pad unit has parametric dimensions and includes two cords as illustrated in Figure 6.

The simulations are all performed assuming linear elasticity, to focus the attention on the mechanical interaction between matrix and cords independently from other geometrical and constitutive nonlinear effects. Despite elastomers and natural rubber might exhibit a nonlinear elastic response, we considered cords made of steel wires – actually used to reinforce rubber tires – and initial perfect bonding at the interfaces, in this way obtaining conditions consistent with both the theory and the operational stress-strain regimes of composites in real cases.

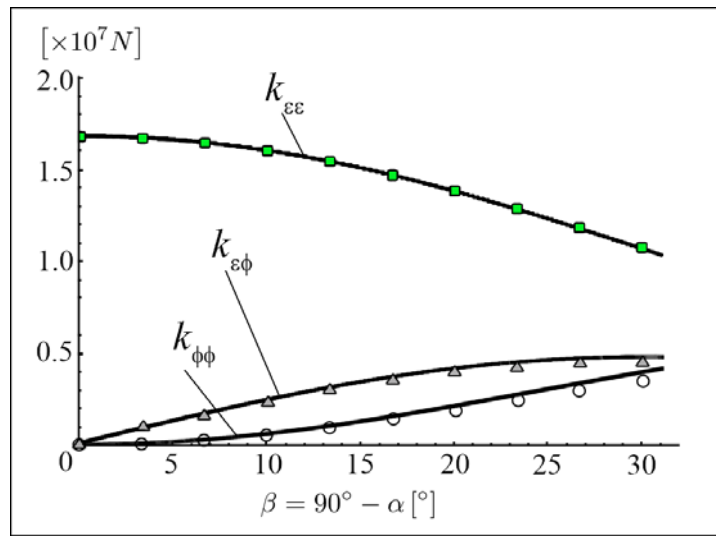

**Figure 5:** Stiffness parameters, including out-of-diagonal coefficients due to the axial-twisting coupling, versus helical angle $\beta$ taken with respect to the strand axis. Comparison between the analytical Argatov model (continuous line) and the cylindrical FE model with a homogenized trigonal material (circular, triangular and square markers, respectively). The (symmetrized) out-of-diagonal stiffness $k_{\varepsilon\phi}$ and the stiffness $k_{\phi\phi}$ have been normalized by $r_h$ and $r_h^2$, respectively. The strand parameters used in the model are $R_c = 1.97$ mm, $R_w = 1.865$ mm, $E_c = E_w = 197.9$ GPa, $v_c = v_w = 0.3, n = 6$.

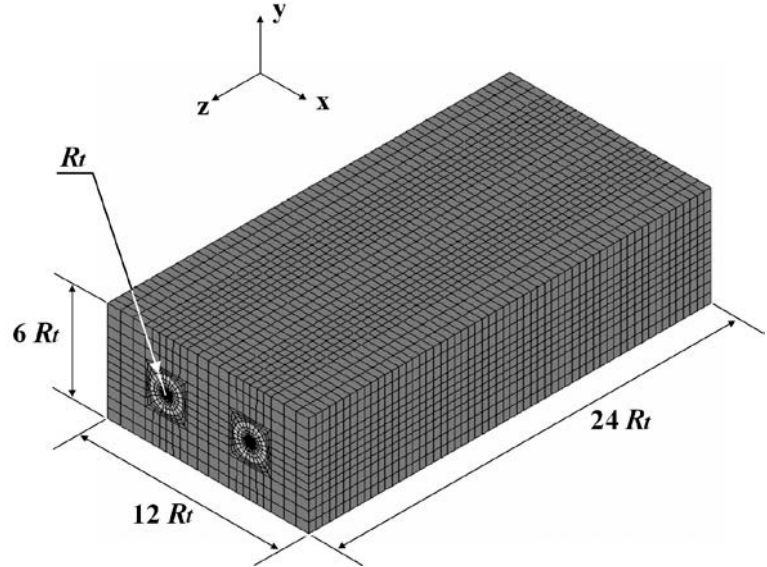

**Figure 6:** FE model of the pad unit of the cord-rubber composite. The rescaling of the mesh size is parameterized with respect to the outer fibre radius.

## *4.2. Stress concentrations and onset of failure*

In order to evaluate how the monoclinic/trigonal behaviour of reinforcing fibres influences the actual distribution of stresses around the fibres and how these can cooperate to trigger the onset of failure phenomena through high stress concentrations in the matrix (e.g. rubber) of a composite, we refer to the 2-unit pad FE model of Figure 7a, subjected to a 1% axial strain along the direction parallel to the cord axes. In order to anticipate the possible propagation of defects in the matrix, a notch is considered at the free sides of the model in correspondence with a so called *cutting plane*, shown in Figure 7a and b. A direct comparison between rubber reinforced with isotropic and with monoclinic/trigonal fibres is then performed by assuming the mechanical properties of the isotropic reinforcement equal to those of the wires in the trigonal cords. Two significant measures of the stress are considered, i.e. the axial stress in the load direction and the von Mises equivalent stress, which is directly related to the deviatoric part of the stress within the material and potentially involved in the onset of failure phenomena. The distribution of these stresses in the rubber matrix, evaluated in correspondence of the cutting plane, are shown in Figure 7c and d. By comparing the results of the two models, one can see that the tensile stresses are essentially coincident but the von Mises stress in the pad with monoclinic/trigonal fibres drastically increases, reaching values up to ten times higher than the ones developing in the pad reinforced with isotropic fibres. These results confirm that critical stresses responsible for possible failure phenomena are strongly underestimated if the chirality effects induced by monoclinic/trigonal twisted fibres are not explicitly taken into account in the composite model. This simplification can lead to large errors in the study of the overall mechanical response of the composite and in the estimation of its mechanical performance. Chiral symmetry is clearly revealed by observing the deformed shape in Figure 7e, in which the mechanical role played by the trigonality of fibres on the overall composite response can be immediately appreciated. Moreover, the inset in Figure 7e shows that the opening mode of the tip is essentially due to the presence of shear stress orthogonal to the load direction [55]. This aspect demonstrates that the shear stresses transmitted to the (rubber) matrix, as predicted by the enriched model reinforced with monoclinic/trigonal fibres, are strictly correlated to the onset of cracking and delamination phenomena actually observed in failure tests performed on rubber/cord composites.

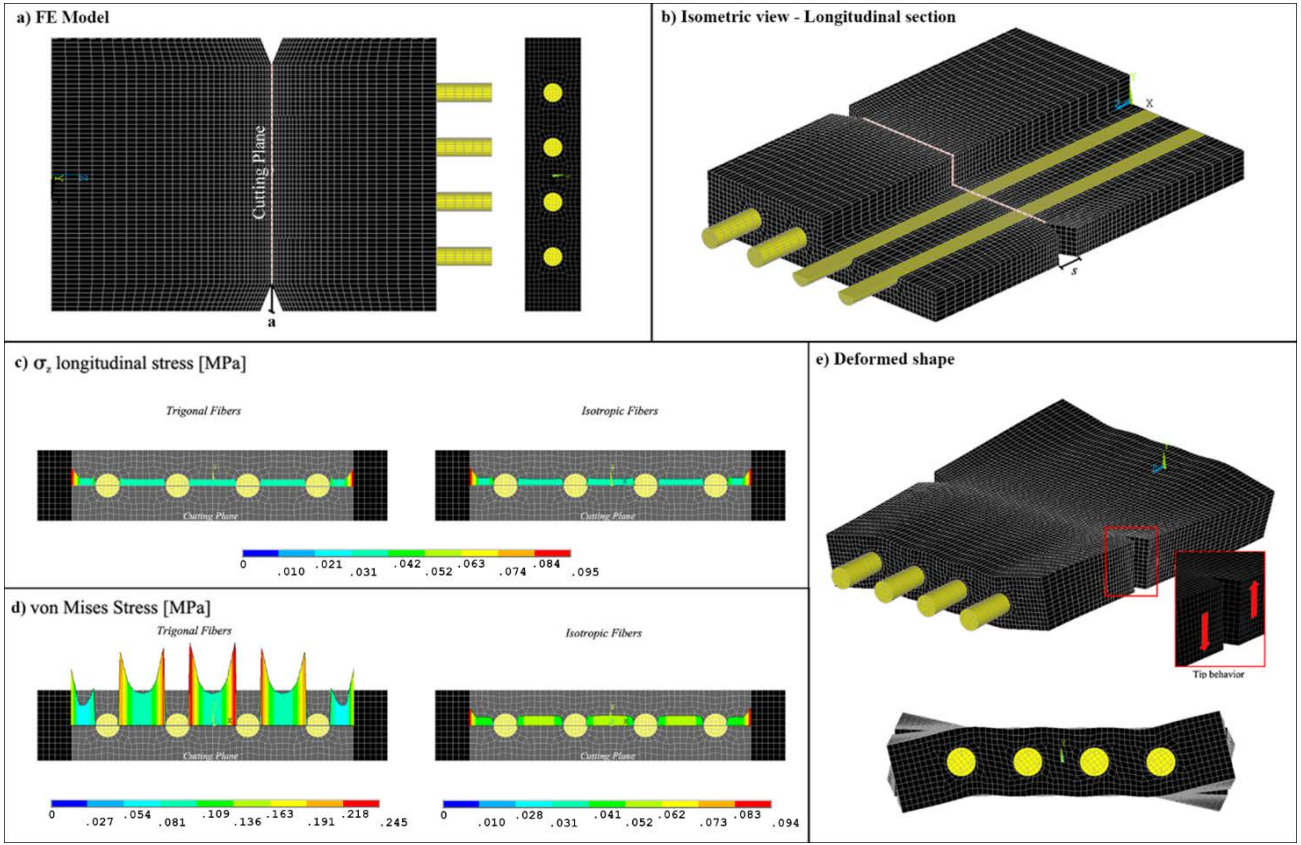

**Figure 7:** a, b) FE model of the rubber pad unit with a series of adjacent parallel cords: a) top and side view; b) isometric view. c) Axial stress in the rubber matrix of the pad respectively for monoclinic/trigonal fibres (*left*) and isotropic fibres (*right*). d) Von Mises stress in the rubber matrix respectively for monoclinic/trigonal fibres (*left*) and isotropic fibres (*right*). e) Deformation of the pad unit subject to 1% of axial strain along the direction parallel to the cord axes. The figure clearly shows the *chiral* symmetry due to the trigonality of the embedded fibres. The tip highlights the presence of shear stress orthogonal to the load direction (displacements are magnified). The geometrical parameters employed in the simulations follow the parameterization shown in Figure 6 with $R_t = 0.5 mm$, and an inter-fibre distance equal to $d_{fib} = 4R_t$, while the constitutive properties used for steel wires and rubber matrix are $E_w = 197.9$ GPa, $E_{rubber} = 3.4$ MPa, $v_w = 0.3$, $v_{rubber} = 0.49$.

Finally, the activation of stresses leading to limit behaviours has been examined by additionally comparing pads reinforced with monoclinic/trigonal fibres with wrapping in the same direction, alternate wrapping directions and reference isotropic fibres (Figure 8). Here, by postulating a perfect elastic-plastic behaviour of the matrix – a scholastic hypothesis introduced with the aim of qualitatively gain insights into the tendency of maximum stress and plastic strains to localize and diffuse in some regions of the matrix at the early stage of the post-elastic behaviour – it is shown that the highest values of the von Mises stresses, leading to the onset of plastic strains as the loads increase, is reached at the rubber-cord interface. In addition, the von Mises stress occurring in the rubber matrix of the two trigonal pads exhibit an almost complementary distribution.

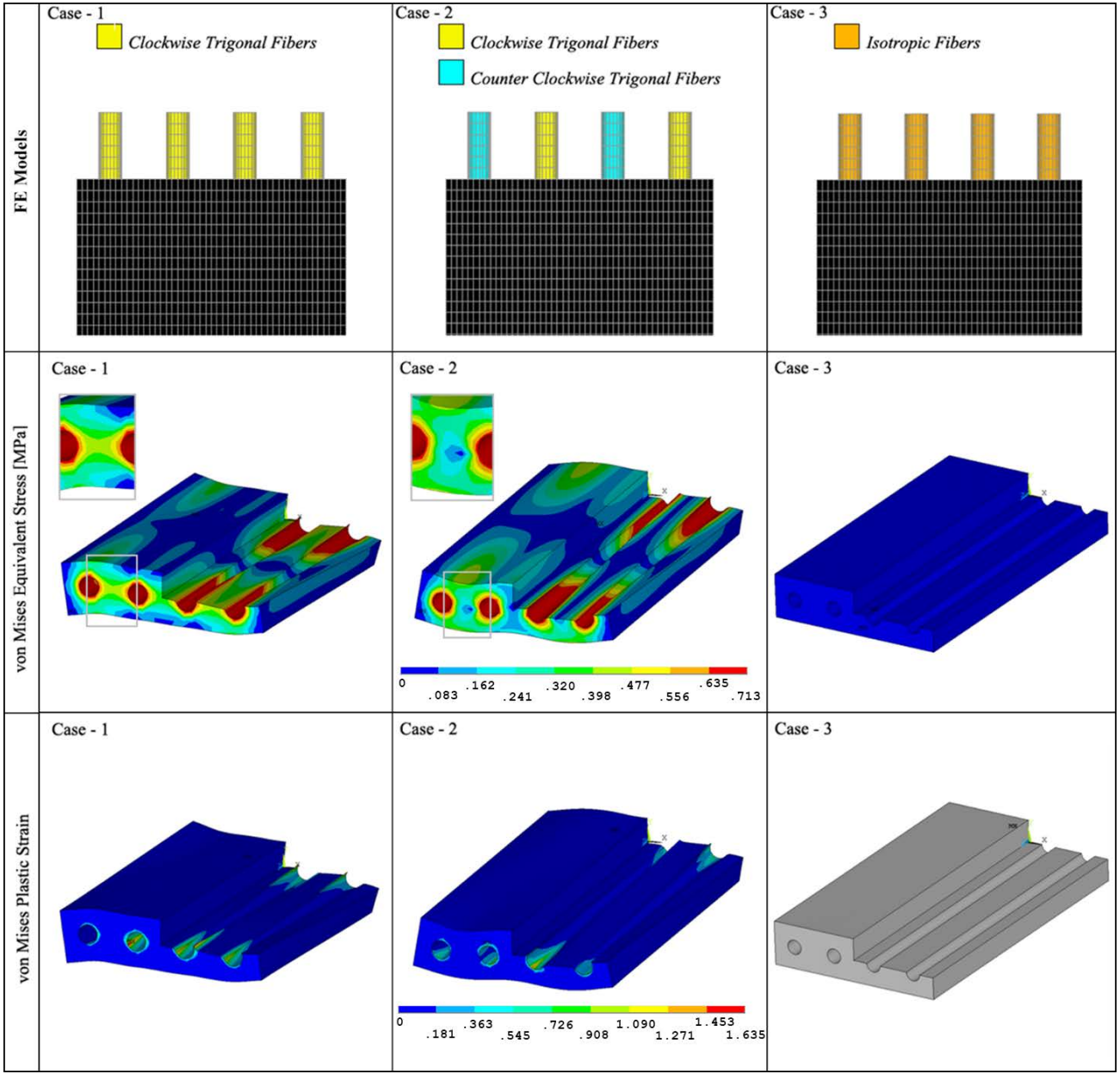

**Figure 8:** Scholastic results obtained by postulating an ideal elastic-plastic behaviour of the matrix for rubber/cord pads with different reinforcements to highlight how chirality of the cords affects the onset and the propagation of residual strain to regions near the interfaces, which would be not involved in case of isotropic fibres. Case 1 (*left*): monoclinic/trigonal fibres with the same wrapping directions; Case 2 (*middle*): monoclinic/trigonal fibres with alternate wrapping directions; Case 3 (*right*): isotropic fibres. The geometrical parameters employed in the simulations follow parameterization shown in Figure 6 with $R_t = 0.5mm$, and an inter-fibre distance equal to $d_{fib} = 4R_t$, while the constitutive properties used for steel wires and rubber matrix are $E_w = 197.9$ GPa, $E_{rubber} = 3.4$ MPa, $v_w = 0.3$, $v_{rubber} = 0.49$.

In Figure 9, the distribution of the shear stress, say $\sigma_{xy}$ over the FE rubber domain for a prescribed helix angle of the cords is reported, highlighting the effect of both different distances and wrapping directions when the pad is subject to axial elongation along the z-axis. Critical shear stress concentrations occur in the case of low inter-fibre distance and cords wrapped in the same sense. Also, the monoclinic/trigonal constitutive behaviour of the fibres highly influences the deformation of the pad. When the composite pad is subject to axial elongation (under free boundary conditions), coupled elongation-torsion occurs in strands and the wound wires, stretched along their helical axes, tend to unwrap. This effect becomes larger as the helix angle decreases. When two cords interact in the same pad, different out of plane bulging-out and/or dishing-in effects can occur as a consequence of the cord wrapping directions. An overall torsion of the pad occurs when the fibres have the same wrapping direction, while, in the case of fibres with opposite wrapping angles, the torsion is avoided but a bending phenomenon takes place in the cross sections normal to the z-axis. The smaller the inter-fibre distance, the higher is the deformation.

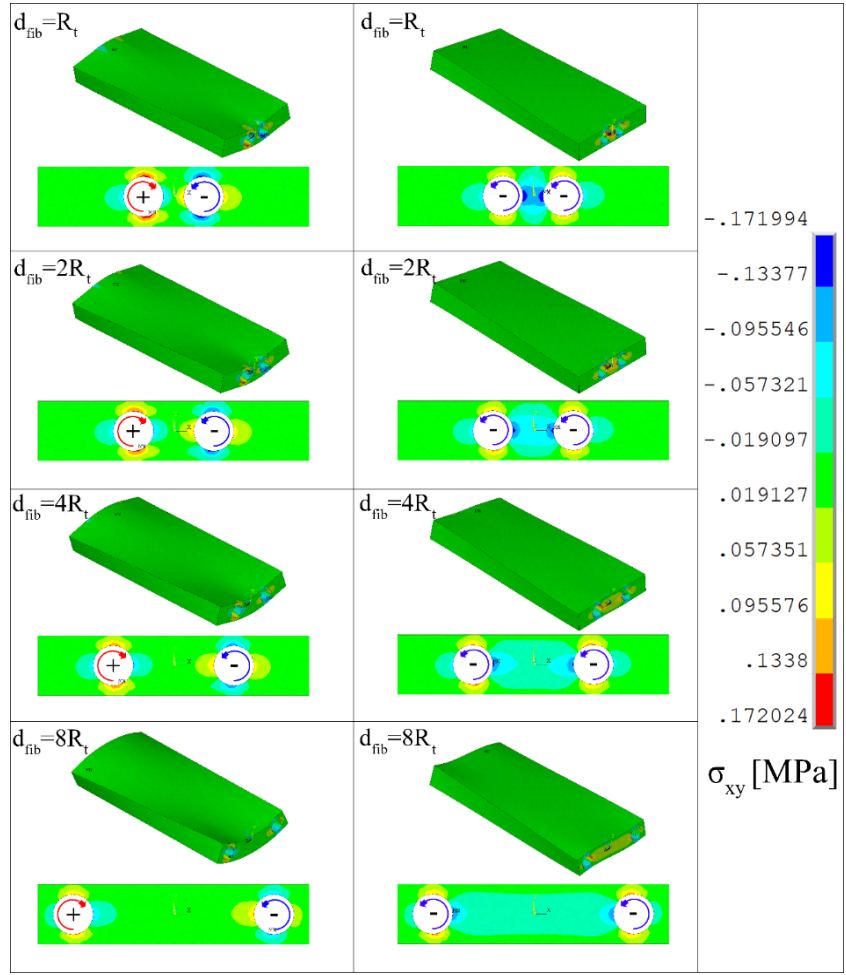

**Figure 9:** Finite element results: contour plot of the shear stress $\sigma_{xy}$ in the pad unit undergoing elongation along the cord axes direction $\varepsilon = 0.01$. The analyses have been performed by considering different cord distances $d_{fib}$. The geometrical parameters employed in the simulations follow the parameterization shown in Figure 6 with $R_t = 0.5 mm$, while the constitutive properties used for the steel wires and rubber matrix are $E_w = 197.9$ GPa, $E_{rubber} = 3.4 MPa$, $v_w = 0.3$, $v_{rubber} = 0.49$.

In Figure 10, the maximum shear stress in the pad unit as a function of the helix angle and cord distances is reported, showing how the trigonality effects decays as the helix angle approaches $\pi/2$.

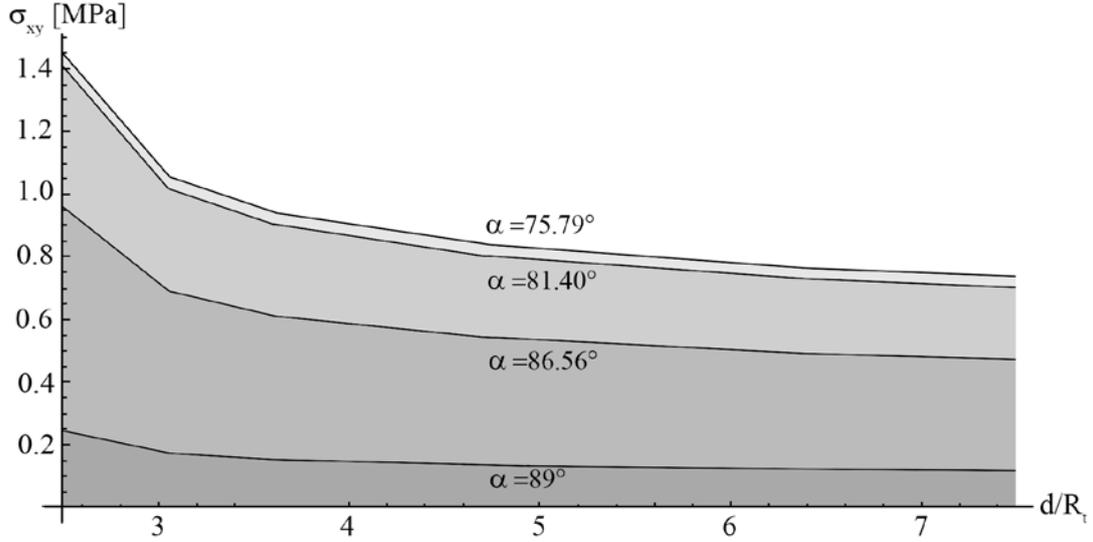

**Figure 10:** Maximum shear stress $\sigma_{xy}$ as a function of the cord distance and helix angle.

## 5  Experimental tests

Experimental mechanical tests were performed on composite specimens fabricated at the European Bridgestone Technical Centre to evaluate how the helicoidal reinforcements affect the response of rubber laminae. In particular, the experimental campaign was aimed at characterizing the mechanical response of three pairs of rubber rectangular specimens – whose length, width and thickness were $L=85\,\text{mm}, w=47\,\text{mm}, t=3\,\text{mm}$ – reinforced with three, five and seven cylindrical fibres, respectively (equally spaced throughout the pad width). Each pair was constituted by a pad with isotropic fibres and one with monoclinic multi-layer strands (Figure 11 a and b), whose detailed mechanical characterization is provided in the work by Fraldi et al. [38]. The mechanical response of these rubber/cord composites was evaluated by means of tensile tests using an INSTRON universal mechanical testing machine mod. 5566 (Instron Corp. Norwood, MA, USA).

In order to measure the in-plane displacements during the experiments, a white square grid with cells of about 2x2 mm$^2$ was stamped on the front face of each specimen (Figure11c). Furthermore, in order to ensure an accurate load transfer at the clamping zone, *ad hoc* clamping systems were designed, able to clamp and grip laterally the rubber samples while keeping the inner cords free (Figure 11d). The tests were carried out under a quasi-static tensile load, and both the elastic and post-elastic evolution of the mechanical response was studied.

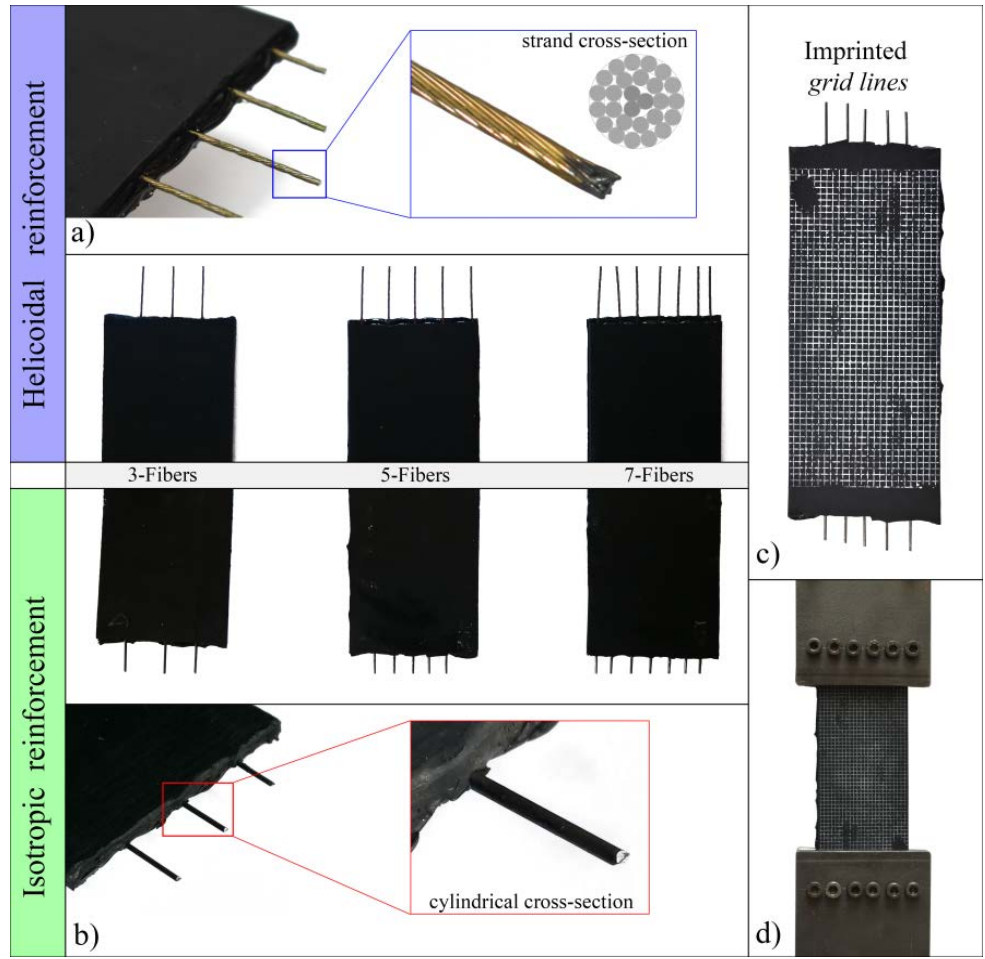

**Figure 11:** Rubber cord specimens manufactured by the European BRIDGESTONE Technical Centre. a) Composites reinforced with helicoidal steel multi-layer strands. b) Composites reinforced with isotropic steel fibres. c) Typical grid lines imprinted on lateral surface of specimens. d) Clamping system.

## 5.1 Analysis of the results of tensile tests

A tensile load was applied longitudinally along the axial fibre direction, by imposing a constant displacement velocity nominally of about $1\,\text{mm/s}$. Results in terms of force-displacement provide important information about the overall mechanical properties of the composites, allowing to gain some insight in the mechanical phenomena occurring at the composite interfaces, as well as in the evolution of the post-elastic phase, during which both failure or ductility features can be evaluated. The obtained force-strain curves are shown in Figure 12. For each test, three zones can be distinguished, corresponding to a first elastic response, followed by the post-elastic behaviour of the composites, including failure events. As highlighted in Figure 12, the elastic response at lower strains, as well as the maximum reaction forces measured, significantly depend on the volume fraction of the reinforcement. Different considerations can be made by examining the post-elastic behaviour of the different reinforced composites, in which the cord/rubber interface plays a crucial rule in the composite strength, as also discussed in the previous Sections.

Two aspects mainly emerge from the results. Comparing pads with the same fraction of fibres, it appears that the peak of the reaction force is always larger in composites with monoclinic fibres, while pads reinforced with isotropic cords show a greater ductility. These features can be effectively correlated to the stress state that develops at the rubber/cord interfaces, which is induced by the anisotropic behaviour of the helicoidal strands, with the larger shear stresses *de facto* leading to premature material failure at lower strains. Focusing on the onset of damaging in pads with helicoidal strands, cracking takes place just ahead the clamps, starting from the central fibres and successively propagating towards the lateral regions of the specimens (Figure 12).

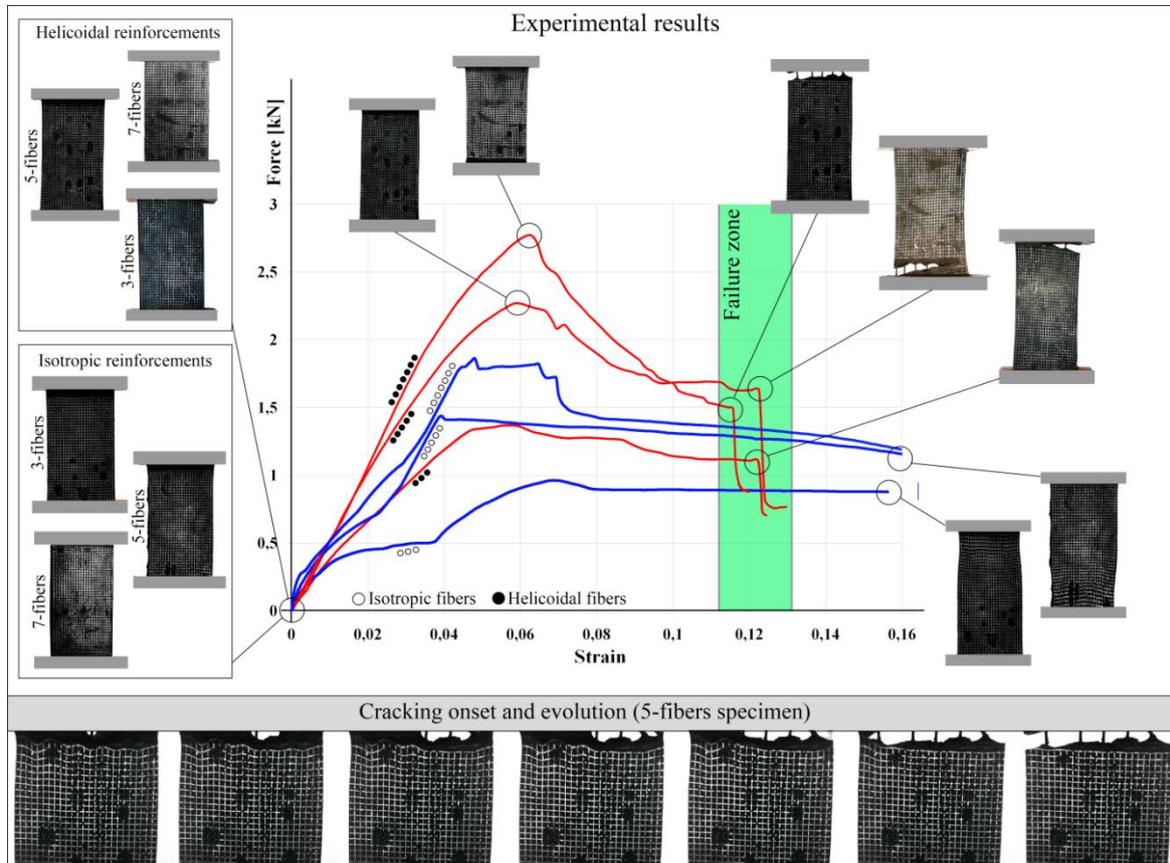

**Figure 12:** Force-strain curves obtained from uniaxial tensile tests in composite pads reinforced with isotropic fibres (blue curves) and with monoclinic/trigonal fibres (red curves).

An ad hoc FE model of the pads was further implemented in order to specifically simulate the performed mechanical tests and to numerically estimate the magnitude of the von Mises stresses involved, as well as of the shear stresses occurring around each fibre (see Figure13a and c). Numerical simulations qualitatively agree with the behaviour of the pads observed for each test, reproducing the in-plane deformation of printed grid lines, which, as experimentally noticed, is mainly governed by the kinematics of the cords within the specimens, the matrix rubber consequently exhibiting the wavy patterns highlighted in Figure13b on its surface. Finally, a comparison between the response of the tested rubber/cord composites with helicoidal fibres and FE simulation including the homogenized equivalent monoclinic cylinders of interest is provided in terms of stiffness in Figure13b.

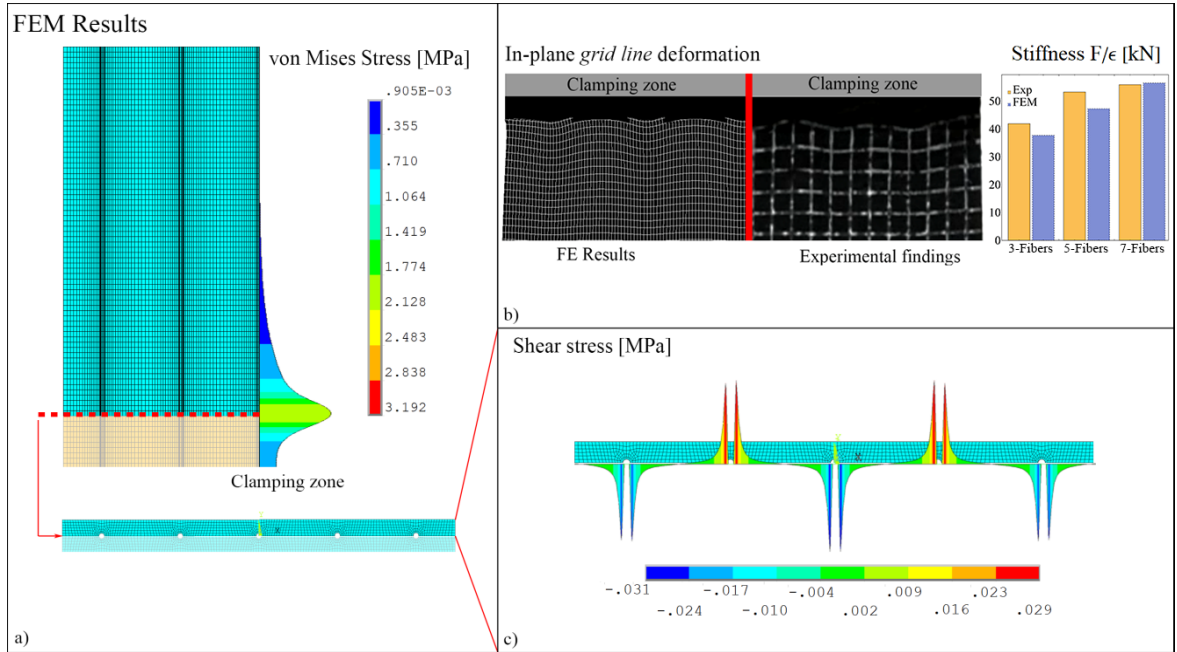

**Figure 13:** Numerical simulations of the mechanical tests. a) Evaluation of the von Mises stress in the clamping zone. b) Comparison between experimental tests and FE simulations of rubber/trigonal-fiber pads: replication of the observed matrix deformation (on the left) and overall stiffness response (on the right). c) Evaluation of the shear stress concentrations at the rubber-cord interfaces.

## 6. Conclusions

We have discussed the effects of stress amplification occurring in composites at the reinforcement-matrix interfaces, and in the neighbouring regions, as a consequence of characteristic stress regimes emerging from the explicit coupling between torsion and elongation in cylindrical strands formed by fibres with helical microstructure. To do this, we first examined the behaviour of the strand as a structure made of a discrete number of wires with helical geometry, and then we derived *ad hoc* closed-form solutions to describe an equivalent monoclinic/trigonal solid cylinder to be used for simulating the reinforcements in composites. The obtained model for homogenized cords was then implemented in FE codes to characterize the mechanical response – in the linear elastic regime and at the onset of crack initiation phenomena – in composites reinforced with helically wound fibres. Without loss of generality, cord-rubber composites were considered, consisting of a rubber matrix in which polymeric or steel reinforcing fibres, characterized by a helical microstructure (i.e. wound filaments) are embedded, these cases being of great interest, for example, in tire applications. Generally, these fibres are spatially variable and are arranged with different orientations and concentrations in order to optimize the mechanical performance of the material under the expected service load and boundary conditions. We found in fact that the distance between cords as well as their orientation largely influence the magnitude of the stresses, leading to concentrations of deviatoric elastic energy in specific regions, for example at cord-rubber and ply-interfaces, where fatigue and aging can give rise to or accelerate the onset of delamination phenomena and crack propagation. As many basic materials in composites, the rubber matrix is particularly vulnerable to the deviatoric part of the stress field and thus to shear stress, due to its incompressibility.

Although both closed-form solutions and numerical results were obtained within the framework of anisotropic linear theory of elasticity (neglecting the intrinsic well-known hyper-elasticity of elastomers due the strong influence of the stiff reinforcing fibres that allow the overall composite response to remain confined to small deformation regimes up to the onset of critical phenomena), the obtained results can be helpful in predicting and explaining some phenomena that could be prodromal to the onset of failure in rubber/cord composites and that have not up to now been completely clarified from a theoretical standpoint.

Evidently, a linear elastic strand model, on which the analytical homogenization procedure is based, does not provide any structural constraint due to inter-wire kinematics [56] as well as disallows stress limitations. It thus neglects the eventual occurrence of plastic phenomena within the helix for the parameter ranges considered in the theoretical predictions, which instead seem to possibly arise also for strains below 1% [57]. However, performing post-elastic analysis is beyond the scope of the present work, which instead aims to reveal the presence of underestimated spurious shear stresses with the related potential risks. The findings and the proposed strategy could be fruitfully employed to implement robust FE-based numerical platforms to analyse the post-elastic response of composites and the stress regimes in the critical regions of the materials, as well as to implement improved design optimization procedures for composites at all scales. In fact, by virtue of the analytically-based formulation, the obtained homogenized cylinders could help to investigate the influence of different helical assemblies in order to find optimal structural arrangements that maximize either the strand or the composite performances [58].

**Funding**

A.C., S.P. and M.F. are supported by Regione Campania under the CIRO (Campania Imaging Infrastructure for Research in Oncology) and SATIN (Therapeutic Strategies against Resistant Cancer) grants and by the Italian Ministry of Education, University and Research (MIUR) under the ARS01-01384-PROSCAN and the PRIN-20177TTP3S grants. A.R.C. also acknowledges support from PON-AIM1849854-1. F.B. is supported by the European Commission H2020 FET Open "Boheme" grant no. 863179. NMP is supported by the European Commission under the Graphene Flagship Core 2 Grant no. 785219 (WP14 "Composites") FET Proactive "Neurofibres" Grant no. 732344 and FET Open "Boheme" grant no. 863179, as well as by the Italian Ministry of Education, University and Research (MIUR) under the "Departments of Excellence" Grant L.232/2016, the ARS01-01384-PROSCAN Grant and the PRIN-20177TTP3S.

**Acknowledgements**

The authors sincerely thank the support from BRIDGESTONE T.C.E. for materials to be tested and Dr. G. Perrella for his contribution in developing some calculations and figures the most part of which have been already reported in a co-authored work (Fraldi et al., JMPS, 2017), cited in the References.

**Appendix A**

**Linear elastostatic problem of a generic *n*-phase monoclinic cylinder**

Quasi-static Cauchy equilibrium equations write in a cylindrical coordinate system $\{r,\vartheta,x_3\}$ as:

$$\sigma_{rr,r}+r^{-1}\sigma_{r\vartheta,\vartheta}+\sigma_{r3,3}+r^{-1}(\sigma_{rr}-\sigma_{\vartheta\vartheta})=0$$
$$\sigma_{r\vartheta,r}+r^{-1}\sigma_{\vartheta\vartheta,\vartheta}+\sigma_{\vartheta 3,3}+2r^{-1}\sigma_{r\vartheta}=0 \qquad (A.1)$$
$$\sigma_{r3,r}+r^{-1}\sigma_{\vartheta 3,\vartheta}+\sigma_{33,3}+r^{-1}\sigma_{r3}=0$$

The comma notation denoting partial derivative operation while the body forces are neglected. These equations can be rewritten in the Ting vector formalism [59], by considering the "emerging" stresses (e.g. tractions on the boundary surface of a single phase) in a way to obtain:

$$(r\,\mathbf{t}_r)_{,r}+(\mathbf{t}_\vartheta)_{,\vartheta}+r\,\mathbf{t}_{3,3}+\mathbf{K}\,\mathbf{t}_\vartheta=\mathbf{0} \qquad (A.2)$$

Where $\mathbf{t}_r=\sigma_{ri}$, $\mathbf{t}_\vartheta=\sigma_{\vartheta i}$, $\mathbf{t}_3=\sigma_{3i}$, $i=\{r,\vartheta,x_3\}$ represent the traction vectors on the surfaces $r=const.$, $\vartheta=const.$, and $x_3=const.$, respectively, and $\mathbf{K}$ is a $3\times 3$ matrix given by

$$\mathbf{K}=\begin{bmatrix}0 & -1 & 0\\ 1 & 0 & 0\\ 0 & 0 & 0\end{bmatrix} \qquad (A.3)$$

Compatibility equations $\mathbf{E}=\mathrm{sym}(\mathbf{u}\otimes\nabla)$ relating strains $\varepsilon_{ij}$ to the displacement $\mathbf{u}=\{u_r,u_\vartheta,u_3\}$ result:

$$\varepsilon_{rr}=u_{r,r},\quad \varepsilon_{\vartheta\vartheta}=r^{-1}(u_{\vartheta,\vartheta}+u_r),\quad \gamma_{r\vartheta}=2\varepsilon_{r\vartheta}=r^{-1}(u_{r,\vartheta}+r\,u_{\vartheta,r}-u_\vartheta)$$
$$\varepsilon_{33}=u_{3,3},\quad \gamma_{r3}=2\varepsilon_{r3}=u_{3,r}+u_{r,3},\quad \gamma_{\vartheta 3}=2\varepsilon_{\vartheta 3}=r^{-1}(u_{3,\vartheta}+r\,u_{\vartheta,3}), \qquad (A.4)$$

By then identifying $(1,2,3)\equiv(r,\vartheta,x_3)$, the stress-strain law $\sigma_{ij}=C_{ijhk}\varepsilon_{hk}$ allows to obtain

$$(\mathbf{t}_r)_j=\sigma_{rj}=C_{1jhk}\varepsilon_{hk}=C_{1j11}\varepsilon_{rr}+C_{1j22}\varepsilon_{\vartheta\vartheta}+C_{1j33}\varepsilon_{33}+C_{1j12}\gamma_{r\vartheta}+C_{1j13}\gamma_{r3}+C_{1j23}\gamma_{\vartheta 3} \quad (A.5)$$

Similar equations hold for $(\mathbf{t}_\vartheta)_j$ and for $(\mathbf{t}_3)_j$. By using equations (A.4), one can obtain:

$$\mathbf{t}_r=\mathbf{Q}_{rr}\,\mathbf{u}_{,r}+r^{-1}\mathbf{Q}_{r\vartheta}(\mathbf{u}_{,\vartheta}+\mathbf{K}\,\mathbf{u})+\mathbf{Q}_{r3}\,\mathbf{u}_{,3}$$
$$\mathbf{t}_\vartheta=\mathbf{Q}_{r\vartheta}^T\mathbf{u}_{,r}+r^{-1}\mathbf{Q}_{\vartheta\vartheta}(\mathbf{u}_{,\vartheta}+\mathbf{K}\,\mathbf{u})+\mathbf{Q}_{\vartheta 3}\,\mathbf{u}_{,3} \qquad (A.6)$$
$$\mathbf{t}_3=\mathbf{Q}_{r3}^T\,\mathbf{u}_{,r}+r^{-1}\mathbf{Q}_{\vartheta 3}^T(\mathbf{u}_{,\vartheta}+\mathbf{K}\,\mathbf{u})+\mathbf{Q}_{33}\mathbf{u}_{,3}$$

With reference to $n$-FGMCs, the linearly elastic constitutive relations (24) can be particularized for each $j-th$ phase. In the Voigt notation, they read:

$$\sigma_\alpha^{(j)}=c_{\alpha\beta}^{(j)}\varepsilon_\beta^{(j)} \qquad (A.7)$$

and the superscript $(j)$ being added to connote the generic $j$-th phase of the $n$-FGMC. Also the matrices $\mathbf{Q}_{hk}$ for each monoclinic layer take the form:

$$\mathbf{Q}_{rr}^{(j)} = \begin{bmatrix} c_{11}^{(j)} & 0 & 0 \\ 0 & c_{66}^{(j)} & c_{56}^{(j)} \\ 0 & c_{56}^{(j)} & c_{55}^{(j)} \end{bmatrix}, \quad \mathbf{Q}_{r\vartheta}^{(j)} = \begin{bmatrix} 0 & c_{12}^{(j)} & c_{14}^{(j)} \\ c_{66}^{(j)} & 0 & 0 \\ c_{56}^{(j)} & 0 & 0 \end{bmatrix}, \quad \mathbf{Q}_{r3}^{(j)} = \begin{bmatrix} 0 & c_{14}^{(j)} & c_{13}^{(j)} \\ c_{56}^{(j)} & 0 & 0 \\ c_{55}^{(j)} & 0 & 0 \end{bmatrix},$$

$$\mathbf{Q}_{\vartheta\vartheta}^{(j)} = \begin{bmatrix} c_{66}^{(j)} & 0 & 0 \\ 0 & c_{22}^{(j)} & c_{24}^{(j)} \\ 0 & c_{24}^{(j)} & c_{44}^{(j)} \end{bmatrix}, \quad \mathbf{Q}_{\vartheta 3}^{(j)} = \begin{bmatrix} c_{56}^{(j)} & 0 & 0 \\ 0 & c_{24}^{(j)} & c_{23}^{(j)} \\ 0 & c_{44}^{(j)} & c_{34}^{(j)} \end{bmatrix}, \quad \mathbf{Q}_{33}^{(j)} = \begin{bmatrix} c_{55}^{(j)} & 0 & 0 \\ 0 & c_{44}^{(j)} & c_{34}^{(j)} \\ 0 & c_{24}^{(j)} & c_{33}^{(j)} \end{bmatrix}$$

(A.8)

Under the hypothesis of axis-symmetry of both the geometry of the object (*n*-FGMC) and of the load conditions, the displacement field is independent from the coordinate $\vartheta$, substitution of (A.6) in (A.2) leads to a differential vector equation in terms of the unknown **u** for the *j*-th layer:

$$\mathbf{Q}_{rr}^{(j)} \mathbf{u}^{(j)}_{,rr} + \mathbf{Q}_{33}^{(j)} \mathbf{u}^{(j)}_{,33} + \left( \mathbf{Q}_{r3}^{(j)} + \left( \mathbf{Q}_{r3}^{(j)} \right)^T \right) \mathbf{u}^{(j)}_{,r3} + r^{-2} \mathbf{K} \mathbf{Q}_{\vartheta\vartheta}^{(j)} \mathbf{K} \mathbf{u}^{(j)} +$$

$$+ r^{-1} \left[ \left( \mathbf{Q}_{r\vartheta}^{(j)} \mathbf{K} + \mathbf{K} \left( \mathbf{Q}_{r\vartheta}^{(j)} \right)^T + \mathbf{Q}_{rr}^{(j)} \right) \mathbf{u}^{(j)}_{,r} + \left( \mathbf{K} \mathbf{Q}_{\vartheta 3}^{(j)} + \left( \mathbf{Q}_{\vartheta 3}^{(j)} \right)^T \mathbf{K} + \mathbf{Q}_{r3}^{(j)} \right) \mathbf{u}^{(j)}_{,3} \right] = \mathbf{0}$$

(A.9)

This can be rearranged in compact form:

$$\mathbf{A}_1^{(j)} \mathbf{u}^{(j)}_{,rr} + \mathbf{A}_2^{(j)} \mathbf{u}^{(j)}_{,33} + \mathbf{A}_3^{(j)} \mathbf{u}^{(j)}_{,r3} + r^{-1} \left[ \mathbf{A}_4^{(j)} \mathbf{u}^{(j)}_{,r} + \mathbf{A}_5^{(j)} \mathbf{u}^{(j)}_{,3} \right] + r^{-2} \mathbf{A}_6^{(j)} \mathbf{u}^{(j)} = \mathbf{0} \quad \text{(A.10)}$$

where

$$\mathbf{A}_1^{(j)} = \mathbf{Q}_{rr}^{(j)}, \quad \mathbf{A}_2^{(j)} = \mathbf{Q}_{33}^{(j)}, \quad \mathbf{A}_3^{(j)} = \mathbf{Q}_{r3}^{(j)} + \left( \mathbf{Q}_{r3}^{(j)} \right)^T, \quad \mathbf{A}_4^{(j)} = \mathbf{Q}_{r\vartheta}^{(j)} \mathbf{K} + \mathbf{K} \left( \mathbf{Q}_{r\vartheta}^{(j)} \right)^T + \mathbf{Q}_{rr}^{(j)},$$

$$\mathbf{A}_5^{(j)} = \mathbf{K} \mathbf{Q}_{\vartheta 3}^{(j)} + \left( \mathbf{Q}_{\vartheta 3}^{(j)} \right)^T \mathbf{K} + \mathbf{Q}_{r3}^{(j)}, \quad \mathbf{A}_6^{(j)} = \mathbf{K} \mathbf{Q}_{\vartheta\vartheta}^{(j)} \mathbf{K},$$

(A.11)

By focusing on a FGMC made of a central *core*, denoted by $(c)$, and *n* arbitrary monoclinic *cladding* phases, the displacement solution for *j*-th generic phase of the (*n*+1)-FGMC is:

$$u_r^{(j)} = u_r^{(j)}(r, x_3), \quad u_\vartheta^{(j)} = u_\vartheta^{(j)}(r, x_3), \quad u_3^{(j)} = u_3^{(j)}(r, x_3), \quad R^{(j-1)} < r < R^{(j)} \quad \text{(A.12)}$$

Applying the compatibility equations (A.4), the stress vector becomes:

$$\begin{bmatrix} \sigma_{rr}^{(j)} \\ \sigma_{\vartheta\vartheta}^{(j)} \\ \sigma_{33}^{(j)} \\ \sigma_{\vartheta 3}^{(j)} \\ \sigma_{r3}^{(j)} \\ \sigma_{r\vartheta}^{(j)} \end{bmatrix} = \begin{bmatrix} c_{11}^{(j)} & c_{12}^{(j)} & c_{13}^{(j)} & c_{14}^{(j)} & 0 & 0 \\ c_{12}^{(j)} & c_{22}^{(j)} & c_{23}^{(j)} & c_{24}^{(j)} & 0 & 0 \\ c_{13}^{(j)} & c_{23}^{(j)} & c_{33}^{(j)} & c_{34}^{(j)} & 0 & 0 \\ c_{14}^{(j)} & c_{24}^{(j)} & c_{34}^{(j)} & c_{44}^{(j)} & 0 & 0 \\ 0 & 0 & 0 & 0 & c_{55}^{(j)} & c_{56}^{(j)} \\ 0 & 0 & 0 & 0 & c_{56}^{(j)} & c_{66}^{(j)} \end{bmatrix} \begin{bmatrix} u_{r,r}^{(j)} \\ u_r^{(j)}/r \\ u_{3,3}^{(j)} \\ u_{\vartheta,3}^{(j)} \\ u_{r,3}^{(j)} + u_{3,r}^{(j)} \\ u_{\vartheta,r}^{(j)} - u_\vartheta^{(j)}/r \end{bmatrix} \quad \text{(A.13)}$$

By requiring the stress components to be independent of $x_3$, the following differential equations can be obtained:

$$u_{r,r3}^{(j)} = 0, \quad u_{r,3}^{(j)} = 0, \quad u_{3,33}^{(j)} = 0, \quad u_{\vartheta,33}^{(j)} = 0, \quad u_{r,33}^{(j)} + u_{3,r3}^{(j)} = 0, \quad u_{\vartheta,r3}^{(j)} + \frac{u_{\vartheta,3}^{(j)}}{r} = 0, \quad (A.14)$$

The displacements solution of equations (A.14) have expression

$$\begin{aligned} u_r^{(j)} &= p_1^{(j)}(r), \\ u_\vartheta^{(j)} &= p_2^{(j)}(r) + \phi^{(j)} r x_3, \\ u_3^{(j)} &= p_3^{(j)}(r) + \varepsilon_0^{(j)} x_3, \end{aligned} \qquad (A.15)$$

where $p_1^{(j)}(r), p_2^{(j)}(r), p_3^{(j)}(r)$ are unknown functions of the sole radial coordinate. By then substituting equations (A.15) and (A.11) into the Navier-Cauchy equations (A.10), the following ODE system is derived:

$$\begin{cases} c_{11}^{(j)}\left(p_{1,r}^{(j)} + r\, p_{1,rr}^{(j)}\right) - c_{22}^{(j)} p_1^{(j)} + \left(c_{13}^{(j)} - c_{23}^{(j)}\right)\varepsilon_0^{(j)} r + \left(2c_{14}^{(j)} - c_{24}^{(j)}\right)\phi^{(j)} r^2 = 0 \\ c_{56}^{(j)}\left(2r\, p_{3,r}^{(j)} + r^2 p_{3,rr}^{(j)}\right) + c_{66}^{(j)}\left(r^2 p_{2,rr}^{(j)} + r\, p_{2,r}^{(j)} - p_2^{(j)}\right) = 0 \\ c_{55}^{(j)}\left(p_{3,r}^{(j)} + r\, p_{3,rr}^{(j)}\right) + c_{56}^{(j)} r\, p_{2,rr}^{(j)} = 0 \end{cases} \qquad (A.16)$$

The function $p_1^{(j)}(r)$ is found by integrating the equation $(A.16)_1$, that is

$$p_1^{(j)}(r) = C_1^{(j)}\left(r^{\lambda^{(j)}} + r^{-\lambda^{(j)}}\right) + i C_2^{(j)}\left(r^{\lambda^{(j)}} - r^{-\lambda^{(j)}}\right) + h_1^{(j)} \varepsilon_0^{(j)} r + h_2^{(j)} \phi^{(j)} r^2 \quad (A.17)$$

where $h_1^{(j)} = \dfrac{c_{23}^{(j)} - c_{13}^{(j)}}{c_{11}^{(j)} - c_{22}^{(j)}}, \quad h_2^{(j)} = \dfrac{c_{24}^{(j)} - 2c_{14}^{(j)}}{4c_{11}^{(j)} - c_{22}^{(j)}}$ and $\lambda^{(j)} = \sqrt{\dfrac{c_{22}^{(j)}}{c_{11}^{(j)}}}$.

The functions $p_2^{(j)}(r)$ and $p_3^{(j)}(r)$ are found by integrating $(A.16)_{2-3}$, i.e.:

$$\begin{aligned} p_2^{(j)}(r) &= h_4^{(j)} C_4^{(j)} - \frac{h_3^{(j)} C_3^{(j)}}{r} \\ p_3^{(j)}(r) &= \frac{h_4^{(j)} C_3^{(j)}}{r} + h_5^{(j)} C_4^{(j)} \log r \end{aligned} \qquad (A.18)$$

where $h_3^{(j)} = \dfrac{c_{55}^{(j)}}{2\left(c_{56}^{(j)2} - c_{55}^{(j)} c_{66}^{(j)}\right)}, \quad h_4^{(j)} = \dfrac{c_{56}^{(j)}}{\left(c_{56}^{(j)2} - c_{55}^{(j)} c_{66}^{(j)}\right)}, \quad h_5^{(j)} = \dfrac{c_{66}^{(j)}}{\left(c_{56}^{(j)2} - c_{55}^{(j)} c_{66}^{(j)}\right)}$.

Then, the displacement solution for a generic monoclinic hollow cylinder is:

$$\begin{aligned} u_r^{(j)} &= C_1^{(j)}\left(r^{\lambda^{(j)}} + r^{-\lambda^{(j)}}\right) + i C_2^{(j)}\left(r^{\lambda^{(j)}} - r^{-\lambda^{(j)}}\right) + h_1^{(j)} \varepsilon_0^{(j)} r + h_2^{(j)} \phi^{(j)} r^2 \\ u_\vartheta^{(j)} &= \phi^{(j)} r x_3 + h_4^{(j)} C_4^{(j)} - \frac{h_3^{(j)} C_3^{(j)}}{r} \\ u_3^{(j)} &= \varepsilon_0^{(j)} x_3 + \frac{h_4^{(j)} C_3^{(j)}}{r} + h_5^{(j)} C_4^{(j)} \log r \end{aligned} \qquad (A.19)$$

The displacement solution for the internal isotropic core phase $(c)$ reduces to equation (31):

$$u_r^{(c)} = C_0^{(c)} r$$
$$u_\vartheta^{(c)} = \phi^{(c)} r x_3 \quad (A.20)$$
$$u_3^{(c)} = \varepsilon_0^{(c)} x_3$$

Constitutive law (A.13) written in case of isotropy case together with solutions (A.20) give the core stresses:

$$\sigma_{rr}^{(c)} = \left(c_{11}^{(c)} + c_{12}^{(c)}\right) C_0^{(c)} + c_{12}^{(c)} \varepsilon_0^{(c)}$$
$$\sigma_{\vartheta\vartheta}^{(c)} = \left(c_{11}^{(c)} + c_{12}^{(c)}\right) C_0^{(c)} + c_{12}^{(c)} \varepsilon_0^{(c)}$$
$$\sigma_{33}^{(c)} = 2 c_{12}^{(c)} C_0^{(c)} + c_{11}^{(c)} \varepsilon_0^{(c)} \quad (A.21)$$
$$\sigma_{\vartheta 3}^{(c)} = \left(c_{11}^{(c)} - c_{12}^{(c)}\right) \phi^{(c)} r$$

For the cylindrically monoclinic outer layers, equations (A.13) and (A.19) return:

$$\begin{bmatrix} \sigma_{rr}^{(j)} \\ \sigma_{\vartheta\vartheta}^{(j)} \\ \sigma_{33}^{(j)} \\ \sigma_{\vartheta 3}^{(j)} \\ \sigma_{r3}^{(j)} \\ \sigma_{r\vartheta}^{(j)} \end{bmatrix} = \begin{bmatrix} \left(k_{11}^{(j)} r^{\lambda^{(j)}-1} + k_{12}^{(j)} r^{-(\lambda^{(j)}+1)}\right) & i\left(k_{13}^{(j)} r^{\lambda^{(j)}-1} + k_{14}^{(j)} r^{-(\lambda^{(j)}+1)}\right) & 0 & 0 & k_{15}^{(j)} & k_{16}^{(j)} r \\ \left(k_{21}^{(j)} r^{\lambda^{(j)}-1} + k_{22}^{(j)} r^{-(\lambda^{(j)}+1)}\right) & i\left(k_{23}^{(j)} r^{\lambda^{(j)}-1} + k_{24}^{(j)} r^{-(\lambda^{(j)}+1)}\right) & 0 & 0 & k_{25}^{(j)} & k_{26}^{(j)} r \\ \left(k_{31}^{(j)} r^{\lambda^{(j)}-1} + k_{32}^{(j)} r^{-(\lambda^{(j)}+1)}\right) & i\left(k_{33}^{(j)} r^{\lambda^{(j)}-1} + k_{34}^{(j)} r^{-(\lambda^{(j)}+1)}\right) & 0 & 0 & k_{35}^{(j)} & k_{36}^{(j)} r \\ \left(k_{41}^{(j)} r^{\lambda^{(j)}-1} + k_{42}^{(j)} r^{-(\lambda^{(j)}+1)}\right) & i\left(k_{43}^{(j)} r^{\lambda^{(j)}-1} + k_{44}^{(j)} r^{-(\lambda^{(j)}+1)}\right) & 0 & 0 & k_{45}^{(j)} & k_{46}^{(j)} r \\ 0 & 0 & 0 & -r^{-1} & 0 & 0 \\ 0 & 0 & -r^{-2} & 0 & 0 & 0 \end{bmatrix} \begin{bmatrix} C_1^{(j)} \\ C_2^{(j)} \\ C_3^{(j)} \\ C_4^{(j)} \\ \varepsilon_0^{(j)} \\ \phi^{(j)} \end{bmatrix}$$

(A.22)

in which

$$\begin{aligned}
&k_{11}^{(j)} = k_{13}^{(j)} = c_{12}^{(j)} + c_{11}^{(j)}\lambda^{(j)}, \quad k_{12}^{(j)} = -k_{14}^{(j)} = c_{12}^{(j)} - c_{11}^{(j)}\lambda^{(j)}, \\
&k_{15}^{(j)} = k_{25}^{(j)} = \frac{c_{23}^{(j)}\left(c_{11}^{(j)} + c_{12}^{(j)}\right) - c_{13}^{(j)}\left(c_{12}^{(j)} + c_{22}^{(j)}\right)}{c_{11}^{(j)} - c_{22}^{(j)}}, \\
&k_{26}^{(j)} = 2k_{16}^{(j)} = \frac{2c_{24}^{(j)}\left(2c_{11}^{(j)} + c_{12}^{(j)}\right) - 2c_{14}^{(j)}\left(2c_{12}^{(j)} + c_{22}^{(j)}\right)}{4c_{11}^{(j)} - c_{22}^{(j)}}, \\
&k_{21}^{(j)} = k_{23}^{(j)} = c_{22}^{(j)} + c_{12}^{(j)}\lambda^{(j)}, \quad k_{22}^{(j)} = -k_{24}^{(j)} = c_{22}^{(j)} - c_{12}^{(j)}\lambda^{(j)}, \\
&k_{31}^{(j)} = k_{33}^{(j)} = c_{23}^{(j)} + c_{13}^{(j)}\lambda^{(j)}, \quad k_{32}^{(j)} = -k_{34}^{(j)} = c_{23}^{(j)} - c_{13}^{(j)}\lambda^{(j)}, \\
&k_{35}^{(j)} = c_{33}^{(j)} - \frac{c_{13}^{(j)2} - c_{23}^{(j)2}}{c_{11}^{(j)} - c_{22}^{(j)}}, \quad k_{36}^{(j)} = c_{34}^{(j)} + \frac{\left(c_{24}^{(j)} - 2c_{14}^{(j)}\right)\left(c_{23}^{(j)} + 2c_{13}^{(j)}\right)}{4c_{11}^{(j)} - c_{22}^{(j)}}, \\
&k_{41}^{(j)} = k_{43}^{(j)} = c_{24}^{(j)} + c_{14}^{(j)}\lambda^{(j)}, \quad k_{42}^{(j)} = -k_{44}^{(j)} = c_{24}^{(j)} - c_{14}^{(j)}\lambda^{(j)}, \\
&k_{45}^{(j)} = c_{34}^{(j)} - \frac{\left(c_{14}^{(j)} + c_{24}^{(j)}\right)\left(c_{13}^{(j)} - c_{23}^{(j)}\right)}{c_{11}^{(j)} - c_{22}^{(j)}}, \quad k_{46}^{(j)} = c_{44}^{(j)} + \frac{c_{24}^{(j)2} - 4c_{14}^{(j)2}}{4c_{11}^{(j)} - c_{22}^{(j)}}
\end{aligned} \quad (A.23)$$

By assuming perfect contact at the cylindrical interfacial boundaries, interface and boundary conditions have to be introduced. The total unknown parameters to be determined are:

$$\begin{aligned}
&C_0^{(c)}, \phi^{(c)}, \varepsilon_0^{(c)} \\
&C_1^{(j)}, C_2^{(j)}, C_3^{(j)}, C_4^{(j)}, \phi^{(j)}, \varepsilon_0^{(j)} \quad j \in \{1, 2, \ldots n\}
\end{aligned} \quad (A.24)$$

where the 3 coefficients in $(A.24)_1$ represent the unknowns of the core, while the $6n$ coefficients in $(A.24)_2$ are the unknowns of each cylindrical hollow phase, the last two coefficients representing the unit angle warping and the axial strain of each layer, respectively. The total number of unknowns is $(6n+3)$, matching the $6n$ interface conditions and 3 external boundary conditions, written at the outer cylindrical surface and at the end basis of the FGMC. In particular, continuity of displacement and stresses at the each interface give:

$$\begin{cases} u_i^{(j)}(r = R^{(j)}) = u_i^{(j+1)}(r = R^{(j)}) \\ \sigma_{ri}^{(j)}(r = R^{(j)}) = \sigma_{ri}^{(j+1)}(r = R^{(j)}) \end{cases} \quad j \in \{0, 1, \ldots, n-1\}, \ i = \{r, \vartheta, x_3\} \quad (A.25)$$

where $R^{(j)}$ is the outer radius of the $j$-th phase, $j = 0$ indicating the core. The respect of the equations (A.25) implies that:

$$\phi^{(c)} = \phi^{(j)} = \phi, \quad \varepsilon_0^{(c)} = \varepsilon_0^{(j)} = \varepsilon_0, \quad C_3^{(j)} = C_4^{(j)} = 0, \quad \forall j \in \{0, 1, \ldots, n-1\}, \quad (A.26)$$

Recalling (A.26), the system (A.25) reduces to:

$$\begin{cases} u_{rr}^{(j)}(r = R^{(j)}) = u_{rr}^{(j+1)}(r = R^{(j)}) \\ \sigma_{rr}^{(j)}(r = R^{(j)}) = \sigma_{rr}^{(j+1)}(r = R^{(j)}) \end{cases} \quad j \in \{0, 1, \ldots, n-1\} \quad (A.27)$$

Equations (A.27) can be written in explicit as

$$\begin{cases}\left(C_1^{(j)}-C_1^{(j+1)}\right)\left(R^{(j)\lambda^{(j)}}+R^{(j)-\lambda^{(j)}}\right)+i\left(C_2^{(j)}-C_2^{(j+1)}\right)\left(R^{(j)\lambda^{(j)}}-R^{(j)-\lambda^{(j)}}\right)+\left(h_1^{(j)}-h_1^{(j+1)}\right)\varepsilon_0\,R^{(j)}\\+\left(h_2^{(j)}-h_2^{(j+1)}\right)\phi R^{(j)2}+R^{(j)}C_0^{(0)}\delta_{0j}=0\\\left(C_1^{(j)}k_{11}^{(i)}-C_1^{(j+1)}k_{11}^{(j+1)}\right)R^{(j)\lambda^{(j)}-1}+\left(C_1^{(j)}k_{12}^{(j)}-C_1^{(j+1)}k_{12}^{(j+1)}\right)R^{(j)-\lambda^{(j)}-1}+\left(k_{15}^{(j)}-k_{15}^{(j+1)}+\dfrac{E\nu\delta_{0j}}{1-\nu-2\nu^2}\right)\varepsilon_0+\\i\left(C_2^{(j)}k_{13}^{(j)}-C_2^{(j+1)}k_{13}^{(j+1)}\right)R^{(j)\lambda^{(j)}-1}+i\left(C_2^{(j)}k_{14}^{(j)}-C_2^{(j+1)}k_{14}^{(j+1)}\right)R^{(j)-\lambda^{(j)}-1}+\left(k_{16}^{(j)}-k_{16}^{(j+1)}\right)\phi R^{(j)}+\dfrac{EC_0\delta_{0j}}{1-\nu-2\nu^2}=0\end{cases}$$

(A.28)

where $\delta_{0j}$ is the Kronecker symbol and the isotropic Lamé moduli of the core $c_{11}^{(c)}=\dfrac{E(1-\nu)}{1-\nu-2\nu^2}$ and $c_{12}^{(c)}=\dfrac{E\nu}{1-\nu-2\nu^2}$ are introdced. Since the external lateral surface of the cylinder is unloaded, the condition $\sigma_{rr}^{(n)}(r=R^{(n)})=0$ gives the equation:

$$C_1^{(n)}\left(k_{11}^{(n)}R^{(n)\lambda^{(j)}-1}+k_{12}^{(n)}R^{(n)-\lambda^{(j)}-1}\right)+iC_2^{(n)}\left(k_{13}^{(n)}R^{(n)\lambda^{(j)}-1}+k_{14}^{(n)}R^{(n)-\lambda^{(j)}-1}\right)+\\+k_{15}^{(n)}\varepsilon_0+k_{16}^{(n)}\phi R^{(n)}=0$$

(A.29)

Without loss of generality, equilibrium equations in the $x_3$ direction at the cylinder basis are written for $x_3=0$:

$$\int_0^{2\pi}\int_0^{R^{(0)}}\sigma_{33}^{(0)}(x_3=0)rdrd\vartheta+\sum_{j=1}^n\int_0^{2\pi}\int_{R^{(j-1)}}^{R^{(j)}}\sigma_{33}^{(j)}(x_3=0)\,rdrd\vartheta=F_3,\\\int_0^{2\pi}\int_0^{R^{(0)}}\sigma_{\vartheta3}^{(0)}(x_3=0)r^2drd\vartheta+\sum_{j=1}^n\int_0^{2\pi}\int_{R^{(j-1)}}^{R^{(j)}}\sigma_{\vartheta3}^{(j)}(x_3=0)\,r^2drd\vartheta=\mathsf{M}_t,$$

(A.30)

where $F_3,\mathsf{M}_t$ are the total axial force and torque moment applied at $x_3=0$. Equations (A.28), (A.29) and (A.30) are conveniently re-arranged in a matrix form. Indeed, we introduce the *loads* vector $\mathbf{L}$

$$\mathbf{L}^T=\{0,0,...,0,F_3,\mathsf{M}_t\}\qquad(\text{A.31})$$

and the *unknowns* vector $\mathbf{X}$

$$\mathbf{X}^T=\{C_0^{(0)},C_1^{(1)},C_2^{(1)},C_1^{(2)},C_2^{(2)},...,C_1^{(i)},C_2^{(i)},...,C_1^{(n)},C_2^{(n)},\varepsilon_0,\phi\}\qquad(\text{A.32})$$

so that the set of equations (A.28), (A.29) and (A.30), reads

$$\mathsf{P}\mathbf{X}=\mathbf{L}\qquad(\text{A.33})$$

where, $\mathsf{P}$ is a $(2n+3)\times(2n+3)$ square matrix whose coefficients depend on both the radii and the elastic moduli of the phases.

In Section 3, the problem is specialized to the case of a core surrounded by a sole cladding layer ($n=1$) and the constants $C_0, C_1$ and $C_2$ derived from the interface and boundary conditions (33)-(35) (that represent the particularization of equations (A.28)-(A.29)) have expression:

$$C_0 = \frac{1}{-12C_a}\left\{\varepsilon 6\left[2c_{13}^I\left(c_{11}^I - c_{12}^I\right)R_w\, r_h + \left(c_{11}^I + c_{12}^I\right)\left(R_w^2 + r_h^2\right)\lambda_c - c_{12}^I R_c^2 \lambda_c\right]+\right.$$
$$\left. -\varphi\, 2c_{14}^I\left(c_{11}^I - c_{12}^I\right)\left(R_c^3 - R_t^3\right)\right\}$$

$$C_1 = \frac{1}{-24 c_{11}^I C_a}\left\{\varepsilon 3 c_{11}^I\left[\left(c_{11}^I - c_{12}^I\right)\left(4c_{13}^I R_w r_h + R_c^2 \lambda_c\right) + 2c_{13}^I R_t^2\left(1 - R_c^2\right)\left(\lambda_c + \mu_c\right)+\right.\right.$$
$$\left.+\left(c_{11}^I + c_{12}^I\right)R_t^2 R_c^2 \lambda_c\right]+$$
$$+\varphi\, c_{14}^I\begin{bmatrix}\left(2\left(c_{11}^I - c_{12}^I\right)\left(\lambda_c + \mu_c\right) + \left(c_{11}^{I\,2} - c_{12}^{I\,2}\right)\right)\left(R_c^2\left(R_c + 2R_t^2 r_h\right) - R_t^3\right)+\\ -2\left(c_{11}^{I\,2} - c_{12}^{I\,2}\right)R_c^2 R_t^2 + 4c_{12}^I R_w^2 R_c^2 R_t^2 +\\ -2c_{12}^I\left(\lambda_c + \mu_c\right)\left(2R_c^5 + 4\left(R_c^2 - 1\right)R_w\left(3r_h + R_w^2\right)\right) + 2c_{11}^I c_{12}^I\left(R_t^3 - R_c^3\right)\end{bmatrix}\right\}$$

$$C_2 = \frac{1}{24 c_{11}^I C_a}\left\{\varepsilon 3 c_{11}^I R_t^2\begin{bmatrix}\left(c_{11}^I - c_{12}^I\right)\left(4c_{13}^I R_w\, r_h + R_c^2 \lambda_c\right)/R_t^2 - \left(c_{11}^I + c_{12}^I\right)R_c^2 \lambda_c +\\ +2c_{13}^I\left(1 + R_c^2\right)\left(\lambda_c + \mu_c\right)\end{bmatrix}+\right.$$
$$\left.-\varphi\, c_{14}^I R_t^2\left[\left(c_{11}^I - c_{12}^I\right)\left|R_t\begin{pmatrix}-R_c^2\left(c_{11}^I + c_{12}^I - 2\left(\lambda_c + \mu_c\right)\right) + \left(c_{11}^I - c_{12}^I + 2\left(\lambda_c + \mu_c\right)\right)\\ \left(R_c + 2R_w - R_c^3/R_t^2\right)\end{pmatrix}+\left(c_{11}^I + c_{12}^I\right)R_c^3\left(c_{11}^I - c_{12}^I + 2\left(\lambda_c + \mu_c\right)\right)\right]\right]\right\}$$

$$C_3 = C_4 = 0 \tag{A.34}$$

where $r_h = R_c + R_w$, and $C_a = \left(c_{11}^{I\,2} - c_{12}^{I\,2}\right)R_w r_h + \left[2\left(c_{11}^I + c_{12}^I\right)R_w r_h + c_{11}^I R_c^2\right]\left(\lambda_c + \mu_c\right)$.

## Appendix B

### Derivation of in-plane trigonal stiffness terms

In order to obtain the explicit expression of the coefficients of the elasticity tensor given in Eq.(45), in what follows we report the essential steps of the analytical approach through which we can derive the homogenized mechanical response of a simple straight strand – modelled as an equivalent cylindrical object – in the plane of its cross section (e.g. the plane whose normal is coaxial to the strand axis). To do this, we assume without loss of generality as isotropic the in plane response of the homogenized trigonal cylinder [46] and consider the elastic problem of a slice of an isotropic hollow cylinder subjected to an external pressure $p$ and whose Lamé moduli $\mu_p$ and $\lambda_p$ are related to Young modulus and Poisson ratio as

$$\lambda_p = \frac{E_p \nu_p}{(1+\nu_p)(1-2\nu_p)} = \lambda_w \gamma^n, \quad \mu_p = \frac{E_p}{2(1+\nu_p)} = \mu_w \gamma^n, \tag{B.1}$$

where $n$ is a parameter to be fixed and $\gamma$ is the volume fraction defined as the ratio between the actual volume occupied by the wires placed around the strand core and the total (nominal) volume of the hollow cylinder, that is

$$\gamma = \frac{m}{4(1 + R_c / R_w) \sin(\alpha)} \quad \text{(B.2)}$$

Here, as in the main text, $R_c$ is the inner radius, $R_t = R_c + 2R_w$ is the outer one and $m$ is the number of wires. The cylindrical reference system $\{r, \vartheta, x_3\}$ is adopted for calculations, with $x_3$ the axis of the cylinder. Under the hypothesis of symmetry (overall symmetrical response of the system in the cross section of the equivalent hallow cylinder) and in absence of body forces and loads applied at both top and bottom ends of the slice, we can assume to work in plane stress.
Therefore, by starting from the classical compatibility, constitutive (stress-strain) and equilibrium equations of the elastic problem and imposing the boundary conditions, one has:

$$\varepsilon_{ij} = \frac{1}{2}\left(\frac{\partial u_i}{\partial x_j} + \frac{\partial u_j}{\partial x_i}\right), \quad \sigma_{ij} = \frac{E_p}{1+\nu_p}\left\{\varepsilon_{ij} + \frac{\nu_p}{1-2\nu_p}\varepsilon_{kk}\delta_{ij}\right\}, \quad \frac{\partial \sigma_{ij}}{\partial x_i} = 0, \quad \forall \boldsymbol{x} \in \Omega \quad \text{(B.3)}$$

B.Cs.: $\{\sigma_{ij} n_i = t_j, \quad \forall \boldsymbol{x} \in \partial\Omega_t; \quad u_i = d_i, \quad \forall \boldsymbol{x} \in \partial\Omega_d\}$

where $t_j$ represents the applied tractions on the boundary $\partial\Omega_t$, $d_i$ are the possible prescribed displacements on $\partial\Omega_d$ and $\partial\Omega = \partial\Omega_t \cup \partial\Omega_d$ is the frontier of the solid domain $\Omega$. The assumptions of axis-symmetry and plane stress mentioned above allow to simplify the problem, which can be reduced to find a displacement $\boldsymbol{u}$ in the form

$$u_r = u^p(r), \quad u_\vartheta = 0, \quad u_3 = \varepsilon_{33}^p x_3 \quad \Rightarrow \quad \varepsilon_{rr}^p = \frac{du^p}{dr}, \quad \varepsilon_{\theta\theta}^p = \frac{u^p}{r}, \quad \varepsilon_{33}^p = const. \quad \text{(B.4)}$$

with $p$ recalling the case at hand in which only uniform pressures are applied at the outmost cylindrical surface of the object. Stress-strain relations for a plane stress state can be thus written as

$$\begin{bmatrix} \sigma_{rr}^p \\ \sigma_{\theta\theta}^p \end{bmatrix} = \frac{E_p}{1-\nu_p^2} \begin{bmatrix} 1 & \nu_p \\ \nu_p & 1 \end{bmatrix} \begin{bmatrix} \varepsilon_{rr}^p \\ \varepsilon_{\theta\theta}^p \end{bmatrix}, \quad \sigma_{33}^p = 0, \quad \varepsilon_{33}^p = -\frac{\nu_p}{E_p}\left(\sigma_{rr}^p + \sigma_{rr}^p\right) \quad \text{(B.5)}$$

While equilibrium equations (B.3) under axis-symmetric assumption write in polar coordinates as

$$\sigma_{rr,r}^p + r^{-1}\left(\sigma_{rr}^p - \sigma_{\vartheta\vartheta}^p\right) = 0 \quad \text{(B.6)}$$

By prescribing a pressure $p$ acting at the outermost surface of the cylinder and a traction-free internal surface, we have that

$$\sigma_{rr}^p(r=R_t)=-p \quad \text{and} \quad \sigma_{rr}^p(r=R_c)=0 \tag{B.7}$$

Finally, by substituting compatibility and constitutive equations (B.4) and (B.5) into (B.6), a second order ODE in terms of the sole radial displacement $u^p(r)$ is obtained, whose solution, after some algebraic manipulations on account of the boundary conditions (B.7), lead to the following expression:

$$u_r^p(r)=\frac{p(1+\nu_p)R_c^2 R_t^2}{E_p(R_t^2-R_c^2)}\left[-\frac{1}{r}-(1-2\nu_p)\frac{R_t^2 r}{R_c^2 R_t^2}\right]-\nu_p r \varepsilon_{33}^p,$$

$$u_\theta^p(r,x_3)=0, \quad u_3^p(x_3)=x_3\,\varepsilon_{33}^p \tag{B.8}$$

the nonzero strains being

$$\varepsilon_{rr}^p=\frac{p(1+\nu_p)R_c^2 R_t^2}{E_p(R_t^2-R_c^2)}\left[\frac{1}{r^2}-(1-2\nu_p)\frac{R_t^2 r}{R_c^2 R_t^2}\right]-\nu_p r\varepsilon_{33}^p,$$

$$\varepsilon_{\theta\theta}^p=\frac{p(1+\nu_p)R_c^2 R_t^2}{E_p(R_t^2-R_c^2)}\left[-\frac{1}{r^2}-(1-2\nu_p)\frac{R_t^2 r}{R_c^2 R_t^2}\right]-\nu_p r\varepsilon_{33}^p, \tag{B.9}$$

$$\varepsilon_{33}^p=-\frac{2\nu_p\, p\, R_t^2}{E_p(R_t^2-R_c^2)},$$

and the stress taking the following form

$$\sigma_{rr}^p=\frac{p\,R_t^2(r^2-R_c^2)}{r^2(R_t^2-R_c^2)}, \quad \sigma_{\theta\theta}^p=\frac{p\,R_t^2(r^2+R_c^2)}{r^2(R_t^2-R_c^2)}, \quad \sigma_{33}^p=0. \tag{B.10}$$

By recalling a classical homogenization approach, we thus consider the overall in-plane behaviour of the isotropic hollow cylinder and write

$$\langle\sigma_{ij}^p\rangle=2\bar{\mu}\langle\varepsilon_{ij}^p\rangle+\bar{\lambda}\langle\varepsilon_{kk}^p\rangle\delta_{ij}, \quad \langle\sigma_{ij}^p\rangle=\frac{1}{\text{Vol}\Omega}\int_\Omega \sigma_{ij}^p\,d\Omega, \quad \langle\varepsilon_{ij}^p\rangle=\frac{1}{\text{Vol}\Omega}\int_\Omega \varepsilon_{ij}^p\,d\Omega \tag{B.11}$$

where $\langle g \rangle$ denotes averaging over the solid domain and $\bar{\mu}$ and $\bar{\lambda}$ represent the homogenized elastic moduli, whose explicit values can be easily determined by using the solutions for stresses and strains found above in the form:

$$\bar{\lambda} = \frac{2R_t^2\left(R_t^2 - R_c^2\right)\lambda_p \mu_p}{\left(R_t^2 + R_c^2\right)\left[3R_c^2\lambda_p + 2\left(R_t^2 + R_c^2\right)\mu_p\right]}, \quad \bar{\mu} = \frac{R_t^2 - R_c^2}{R_t^2 + R_c^2}\mu_p. \tag{B.12}$$

By finally substituting the penalization laws for the stiffness terms (B.1) in the homogenized moduli (B.12), we can find the explicit expressions of the in-plane stiffness terms for the equivalent trigonal cylinder already given in Eq.(45) as follows

$$c_{11}^1 = \gamma^n \frac{16\, r_h\, R_w\, \mu_w \left[R_c^2\left(\lambda_w + \mu_w\right) + R_c R_w\left(\lambda_w + 2\mu_w\right) + R_w^2\left(\lambda_w + 2\mu_w\right)\right]}{\left(R_w^2 + r_h^2\right)\left[8\mu_w R_w r_h + R_c^2\left(3\lambda_w + 4\mu_w\right)\right]},$$

$$c_{12}^1 = \gamma^n \frac{4\, r_h\, R_w\, R_t^2\, \lambda_w\, \mu_w}{\left(R_w^2 + r_h^2\right)\left[8\mu_w R_w r_h + R_c^2\left(3\lambda_w + 4\mu_w\right)\right]}, \quad c_{66}^1 = \gamma^n \frac{2\, r_h\, R_w}{R_w^2 + r_h^2}\mu_w. \tag{A. 12}$$